\documentclass[sigconf,natbib=true, review=false, anonymous=False,printacmref=false]{acmart} %
\renewcommand\footnotetextcopyrightpermission[1]{}

\newcommand\blfootnote[1]{%
  \begingroup
  \renewcommand\thefootnote{}\footnote{#1}%
  \addtocounter{footnote}{-1}%
  \endgroup
}

\usepackage{lipsum}

\usepackage[utf8]{inputenc}
\usepackage[show]{chato-notes}
\usepackage{booktabs} %
\usepackage{amsmath}
\usepackage{multirow}
\usepackage{placeins}
\usepackage{wrapfig}
\usepackage{subfig}
\usepackage{makecell}
\usepackage{enumitem}
\usepackage{algorithm}
\usepackage{algpseudocode}
\usepackage{marginnote}

\usepackage{pgfplots}
\DeclareUnicodeCharacter{2212}{−}
\usepgfplotslibrary{groupplots,dateplot}
\pgfplotsset{compat=newest}

\graphicspath{ {./graphics/} }
\usepackage{amsmath}
\DeclareMathOperator*{\argmax}{arg\,max}

\title{Aligning GPTRec with Beyond-Accuracy Goals  with Reinforcement Learning}

\copyrightyear{2024}
\acmYear{2024}
\setcopyright{acmlicensed}\acmConference[GenRec@WWW '24]{The 2nd Workshop on Recommendation with Generative Models}{May 13--17, 2024}{Singapore, Singapore}
\acmBooktitle{The 2nd Workshop on Recommendation with Generative Models (GenRec@WWW '24), May 13--17, 2024, Singapore, Singapore}

\author{Aleksandr V. Petrov}
\affiliation{%
  \institution{University of Glasgow} \country{United Kingdom}}

\email{a.petrov.1@research.gla.ac.uk}

\author{Craig Macdonald}
\affiliation{%
  \institution{University of Glasgow} \country{United Kingdom}}
\email{craig.macdonald@glasgow.ac.uk}

\newcommand{\www}[1]{\textcolor{black}{#1}}
\newcommand{\slw}[1]{\textcolor{black}{#1}}
\newcommand{\sasha}[1]{\textcolor{black}{#1}}
\newcommand{\sdd}[1]{\textcolor{black}{#1}}
\newcommand{\sdm}[1]{\textcolor{black}{#1}}
\newcommand{\sj}[1]{\textcolor{black}{#1}}
\newcommand{\sjj}[1]{\textcolor{black}{#1}}

\newcommand{\sw}[1]{\textcolor{black}{#1}}

\newcommand{\pageenlarge}[1]{\marginnote{}\enlargethispage{#1\baselineskip}}

\begin{document}
\begin{abstract}
\www{\sdm{Sequential recommender models work with} chronologically ordered sequences of interactions, to predict the next interaction in the sequence. Recently, adaptations of Transformer models, such as BERT4Rec and SASRec, \sasha{have} achieved state-of-the-art performance in the next-item prediction task according to accuracy-based metrics, such as NDCG. These models treat items as tokens, and then utilise a score-and-rank approach (\sasha{\emph{Top-K}} strategy), where the model first computes item scores and then ranks \sdm{them} according to this score. While this approach works well for accuracy-based metrics, it \sdm{is hard to use it} for optimising more complex \sw{beyond-accuracy metrics} such as diversity. \sw{Recently, the GPTRec model, which uses a different} (\sasha{\emph{Next-K} strategy) has been proposed as an alternative to the Top-K models}.  In contrast with traditional Top-K recommendations, \sw{Next-K generates recommendations} item-by-item and, therefore, can account for complex item-to-item interdependencies important for the beyond-accuracy measures. \sw{However, the original GPTRec paper focused only on accuracy in experiments, and did not address how to optimise the model for complex beyond-accuracy metrics.} 
 \sjj{Indeed, \sw{training GPTRec for beyond-accuracy goals is challenging, because} the interaction training data available for training recommender systems is typically not aligned with beyond-accuracy recommendation goals.} To solve the misalignment problem, we train GPTRec using a 2-stage approach: in the first stage, we use a \sjj{teacher-student approach} to train GPTRec, mimicking the behaviour of traditional Top-K models; at the second stage, we use \sw{Reinforcement Learning to align the model for beyond-accuracy goals.} \sjj{Our proposed training scheme}, in principle, can \sjj{align} the model \sjj{with} any measurable recommendation metric. In particular, we experiment with increasing recommendation diversity and reducing popularity bias. \sasha{Our experiments on two datasets show that in 3 out of 4 cases, GPTRec's Next-K generation approach offers a better tradeoff between accuracy and secondary metrics than classic greedy re-ranking techniques. For example, when compared to BERT4Rec, it can simultaneously improve NDCG@10 by 8.8\% while having 8.6\% lower popularity bias when fine-tuned to reduce popularity bias.}
 \pageenlarge{3}
}
\end{abstract}

\maketitle

\vspace{-0.3\baselineskip}
\section{Introduction} \label{sec:intro}
\vspace{-0.7\baselineskip}

\blfootnote{
Accepted for presentation at the 2nd Workshop on Recommendation with Generative Models, in conjunction with the ACM Web Conference 2024,
May 13 -- 17, 2024,  \\
Singapore, Singapore.\\
\copyright Copyright held by the authors}

\looseness -1 \sw{Sequential recommendation models utilise ordered sequences of user-to-item interactions to produce recommendations.} \www{\sw{State-of-the-art} sequential models usually employ a score-and-rank approach to generate recommendations (\sw{known as} the {\em Top-K strategy}~\cite{GPTRec}). The problem with this method is that items are scored independently, and similar items are likely to have similar scores, limiting the method's ability to optimise metrics beyond accuracy. \sw{For example, many researchers argue that recommendation diversity is one of the key components of good recommender systems~\cite{kunaver2017diversity,antikaciogluNewSystemWideDiversity2019a, zhouSolvingApparentDiversityaccuracy2010}. However, under the Top-K strategy, the models' outputs are likely to be dominated by similar types of items, compromising the user experience.} \sw{A recent paper~\cite{GPTRec} proposed GPTRec model, which uses an alternative} \emph{Next-K strategy}, where recommendations are generated item-by-item. In that case, when the model generates a recommendation in position $i$, it already knows what items are recommended in position $1..i-1$, and may adjust the result accordingly. \www{The main challenge with the Next-K strategy is model training. Indeed, standard supervised learning algorithms do not work well for training a generative model, as they require a source of \emph{good} recommendations, which isn't typically available. For example, to train a model to generate diverse and relevant recommendations, we would need a training set of diverse and relevant recommendations for a large number of users. However, to obtain such a set, we would need a recommendation model with at least the same performance as the generative model we intend to train, which we assume is unavailable \sjj{-- indeed, the only available data training data in a typical recommendation scenario is historic user-item interactions. A similar problem exists in the training of language models, in that the available training data is not aligned with the desired task. However, this problem has recently been addressed by the Reinforcement Learning With Human Feedback (RLHF) approach~\cite{ouyangTrainingLanguageModels2022}.} 
\sjj{Inspired by RLHF,} we propose a reinforcement learning-based approach, where we sample recommendations from the current version of the recommendation model ("Actor"), use a separate model to estimate the quality of produced recommendations ("Critic"), and then use the Critic's output to improve the main recommendation model. This approach \sasha{can optimise} for almost any metric measured on a full recommendation list. In particular, we apply this approach to optimise recommendation diversity and reduce popularity bias.}}
\pageenlarge{3}

\looseness -1 \www{\sw{In our proposed 2-stage training scheme, at}  the pre-training stage, the model learns to mimic the teacher model, and at the fine-tuning stage, we align it with beyond-accuracy goals using Reinforcement Learning}. We use the pre-training/fine-tuning approach with GPTRec on two datasets, namely \sasha{MovieLens-1M}~\cite{harperMovieLensDatasetsHistory2015} and \sasha{Steam-2M} (a \sasha{smaller} version of the Steam~\cite{SASRec} dataset). Our experiments show that GPTRec in generative mode matches the performance of state-of-the-art BERT4Rec when tuned for accuracy only. However, when the model is tuned for more \sasha{intricate} goals, such as increasing diversity or decreasing popularity bias, GPTRec provides a better tradeoff between accuracy and one of these goals, compared to applying BERT4Rec with greedy re-ranking techniques such as using Maximum Marginal Relevance (MMR)~\cite{carbonellUseMMRDiversitybased1998b} for improving diversity. \sdm{For example, on Steam-2M, while exhibiting the same amount of diversity (measured by the Intra-List Distance (ILD)~\cite{antikaciogluNewSystemWideDiversity2019a}), GPTRec achieves 10\% higher NDCG@10 than BERT4Rec with MMR.}

\sjj{Following~\cite{deffayetGenerativeSlateRecommendation2023}, for practical reasons, we limit the scope of this paper to datasets with a few thousand items in the catalogue. While we acknowledge the dataset size as a limitation of this research, we argue that there are many industrial applications where the catalogue size does not exceed this size. Examples of such applications include but are not limited to single-brand online shops (e.g. apple.com), news agency websites (it only makes sense to recommend news no older than a few days old), genre recommendation in music (where the system recommends specific genres instead of concrete songs) and many others.}
\sdd{In practice, it is hard to train transformer-based models for recommendations with more than a few hundred thousand items due to memory and computational constraints~\cite{PetrovRSS22}. Moreover, in our case, we have to keep several copies of the model in memory as we use multiple asynchronous processes for training; therefore, we limit the scope of this research to a smaller number of items} to be able to experiment quickly on limited hardware; Furthermore, \sdd{the success of InstructGPT~\cite{ouyangTrainingLanguageModels2022}, which used a similar reinforcement learning approach to train a model with more than 100 billion parameters} allows us to argue that the same approach will work on a much larger scale, but that requires access to a large amount of industry-grade GPUs, which is beyond our \sdd{available} budget\footnote{\sjj{While in this work, we experiment with the datasets with a moderate number of items, we note that scaling is possible by using sub-item representation. However, scaling transformer models for recommendation with large catalogues is an orthogonal research direction, which  \sw{has been addressed in several recent} works~\cite{GPTRec,petrovRecJPQTrainingLargeCatalogue2024, petrovGSASRecReducingOverconfidence2023}. In industrial settings, these limitations can also be mitigated using engineering solutions, such as storing the model in shared memory, but these optimisations are beyond the scope of this paper.}}. 

\sasha{
In short, we summarise the contributions of this paper as follows: 
    (i) \www{we propose a supervised pre-training/reinforcement fine-tuning approach for training \sasha{generative} recommender models for complex recommendation goals; }
    (ii) we show that \sasha{when} optimised for accuracy only, it can match state-of-the-art transformer-based models, such as BERT4Rec while generating recommendations using the Next-K technique;}
    (iii) \www{we show that GPTRec \sasha{can be optimised} for complex recommendation goals, such as increased diversity and decreased popularity bias,  and \sasha{an optimised version of GPTRec provides} a better \sasha{tradeoff between accuracy and secondary metrics} than state-of-the-art BERT4Rec with greedy re-ranking.}

\sasha{The rest of the paper is organised as follows: \sw{Section~\ref{sec:related} covers related work on alignment in language models and sequential recommendation models}.  Section~\ref{sec:gptrec_training} describes the \sw{proposed} two-stage training process for GPTRec; Sections~\ref{sec:exp_setup}~and~\ref{sec:results} contain the experimental evaluation of GPTRec; Section~\ref{sec:conclusion} contains final remarks.} 

\pageenlarge{3}
\vspace{-0.7\baselineskip}
\section{Related work} \label{sec:related}
\sw{In this section we provide a brief overview of related work on alignment in language models, as well as describe GPTRec~\cite{GPTRec} model, which we use in our experiments.}

\vspace{-0.5\baselineskip}
\subsection{\sj{The Misalignment Problem and Reinforce- ment Learning with Human Feedback}} \label{ssec:misalignment}
\sj{One of the big problems of state-of-the-art generative language models is the problem of \emph{misalignment} between training data and the actual desired outcome. A typical desired use-case for generative models is to work in a "chat-bot" mode and follow the instructions of the user. However, the training data for such interactive communication and instruction-following is limited and very expensive to gather. In contrast, "regular" text data is abundant and practically unlimited in the modern days: texts can be crawled from websites, books, public datasets, etc. Earlier generative models, such as GPT-2~\cite{gpt2} and T5~\cite{raffelExploringLimitsTransfer2020}, were trained only using "regular" training data and therefore had limited use for building interactive applications.}

\sj{
Notably, Oyang et al.~\cite{ouyangTrainingLanguageModels2022} proposed a solution for the misalignment problem using the idea of "Reinforcement Learning with Human Feedback" (RLHF). The authors proposed a three-step solution for the problem, \sw{which we now briefly describe step-by-step:} 
     (Step 1) train a language model for the next token prediction task using a "regular" training dataset and fine-tune it using a small amount of available "high-quality" dialogue data that was generated with human labellers;
     (Step 2) use the model from Step (1) to generate different possible outputs for the same model input and ask the labellers to compare generated outputs. These labels are then used to train a \emph{reward} model, which essentially associates a numerical value with the generated output: the larger the reward, the better the output, according to the model; 
     (Step 3) use a reinforcement learning approach from step (1) to fine-tune the language model using the Proximal Policy Optimisation~\cite{schulmanProximalPolicyOptimization2017a} algorithm. 
}

\pageenlarge{3}
\sj{
    We argue that a similar misalignment problem arises in the recommender systems when the recommendation objectives include beyond-accuracy components, such as diversity. Indeed, sequential recommendation datasets typically contain a large number of user-item interaction sequences that can be used for training sequential recommendation models, such as BERT4Rec or SASRec. \sw{Hence}, as the main goal of these models is to predict the next item in the sequence of interactions, the training data is \emph{aligned}, with the target goal of accurately predicting the next item, and these models can achieve state-of-the-art results when trained appropriately using different training objectives, such as sequence shifting, item masking or recency sampling of sequences. However, the datasets typically do not contain "perfect" model outputs for beyond-accuracy goals, such as diversity, and, therefore, supervised training for such goals is challenging due to data scarcity. Usually, there is no source of high-quality training data for beyond-accuracy metrics; the typical solution involves training a model for accuracy (for which training data is abundant) and then re-rank using a heuristic approach, such as MMR~\cite{carbonellUseMMRDiversitybased1998b} or Serendipity Oriented Greedy~\cite{kotkovHowDoesSerendipity2020}. While these heuristics somewhat allow to mitigate the misalignment problem, there is no evidence that such heuristics lead to a solution that may be close to optimal.}
\sj{
    Therefore, inspired by the success of the RLHF for language models, we propose to apply a similar approach to recommender systems. The difference between our approach and the RLHF  is that we assume that the target quality metric is known and measurable (in our experiments, we use a linear combination between accuracy (measured by NDCG) and a secondary objective, such as diversity (for example, measured by Intra-List-Distance~\cite{antikaciogluNewSystemWideDiversity2019a}). This simplifies our approach, as we do not need human labellers to generate high-quality recommendations and compare generated recommendations.} \sjj{Instead, we directly use a known quality measure as a reward and train the model to maximise it using reinforcement learning.}%
    
\subsection{Transformers for sequential recommendation} \label{ssec:gptrec}
\sw{Sequential Recommendation is usually cast as the \emph{next item prediction task}. Formally, given the sequence of historical user-item interactions $\{h_1, h_2...h_n\}$, the goal of the recommender system is to predict the next item $h_{n+1}$. This problem is similar to the \emph{next token prediction} task common in language models, and therefore, adaptations of language models are frequently used for sequential recommendation. In particular, Transformer~\cite{Transformer} based model, such as SASRec~\cite{SASRec}, BERT4Rec~\cite{BERT4Rec} and gSASRec~\cite{petrovGSASRecReducingOverconfidence2023} achieved state-of-the-art results for sequential recommendation. These models employ the Top-K recommendation strategy, which, as we argue in~\ref{sec:intro}, is problematic for the beyond-accuracy metrics.}

\sw{\looseness -1 A recent paper~\cite{GPTRec} proposed  GPTRec, a generative sequential recommendation model that utilises the GPT-2 architecture~\cite{gpt2}, which in turn is based on the Decoder part of the Transformer model~\cite{Transformer}.  In contrast with most existing Transformer-based models, such as BERT4Rec, GPTRec uses the Next-K recommendation approach, i.e. it produces recommendations iteratively (or \emph{autoregressively}, i.e. item-by-item) and, therefore, potentially is more suitable for beyond-accuracy optimisation.}

\looseness -1 \sj{Although GPTRec does not directly use pre-trained weights of its backbone GPT-2 model, it still shares many similarities with GPT. In particular, as we argued in Section~\ref{ssec:misalignment}, there is a misalignment problem: the data available for training usually is not aligned well with the desired recommendation objectives.} 
\sw{In the next section, we describe how we solve this problem using the 2-stage pre-training/fine-tuning approach.}

\section{Training GPTRec for beyond-accuracy goals} \label{sec:gptrec_training}

\sdd{As discussed in  Section ~\ref{ssec:gptrec}, training a version of GPTRec model for Next-K generation \sjj{that is aligned with the beyond-accuracy goals is a challenging task, as training high-quality training samples are rarely available}. In this section, we describe a two-stage process that is capable of addressing this challenge. To do this, we \sdm{first describe the training objectives} in Section~\ref{ssec:problem_formalisation}. We then describe the 2-stage pre-training/fine-tuning approach we use to train the model in Section~\ref{ssec:pre_train_fine_tune}. Finally, \sdm{we describe an efficient process decomposition} for the reinforcement learning fine-tuning in \sdm{Section~\ref{ssec:rl_decompostion}}}. %

\pageenlarge{3}
\subsection{\slw{Training Objectives}} \label{ssec:problem_formalisation}
\slw{
    \sdd{In this section, we formalise the optimisation problem for training GPTRec \sw{for beyond-accuracy goals}}. Consider a sequence of user-item interactions $h=\{h_1, h_2.. h_n\}; h_i \in I$, where $I$ is the catalogue of all available items . The goal of a recommender system is to \emph{generate} a list of recommendations $g=\{g_1, g_2 ... g_K\}; g_i \in I$, which maximises \sdm{an} \emph{effectiveness measure} $R(h, g)$. $R$ may have a complex structure and depend not only on accuracy but also on item popularity,  diversity, etc. Note that $R$ depends on the whole recommendation list, and therefore, it can not be optimised using \sdd{a} pointwise loss functions, such as Binary Cross-Entropy. It also is not required to be differentiable, limiting our ability \sdd{to optimise $R$} using gradient-based methods. Therefore, instead of relying on standard supervised learning, we propose to optimise it using a reinforcement learning approach, which we describe in Section~\ref{ssec:pre_train_fine_tune}}.  

\looseness -1 \slw { 
    Without loss of generality, we can say that metric $R$ can be decomposed into the sum of \emph{immediate rewards} (\sdd{a part of overall effectiveness measure $R$, which can be calculated after generating each individual item in the recommendation list}):
    \begin{align}
        R(h, g) = \sum_{i=1}^{K} \mathrm{r}
(h, g_1, g_2, ... g_i)  =\sum_{i=1}^{K} \mathrm{r}(h, g^{(i)}) \label{eq:immediate_rewards}
    \end{align}
    where $g^{(i)}$ denotes a partially generated list of recommendations up to position $i$:  $g^{(i)} = \{g_1, g_2, .. g_{i}\}$. 
}

\slw{
    Even if the metric is only meaningful over the whole list of items, we can say \sdd{that the immediate reward is 0 for each step, except for the last item in the list; this is also known as \emph{delayed reward}}:
    \begin{align}
    \mathrm{r}(h, g^{(i)}) = 
    \begin{cases}
                0;  i \in 1 .. K-1 \\
                R(h, g); i = K
    \end{cases}  
\end{align}}

\pageenlarge{3}
\sdd{
  However, in many cases, the immediate reward may have a non-zero value even for a partially generated list. For example, if accuracy is part of the overall metric, the immediate reward may be positive when the model generates a relevant item; for decreasing popularity bias, the immediate reward may be negative when generating a popular item. Optimising the model for immediate reward is easier, as it requires the model to do less planning, which is generally a challenging task~\cite{mnihPlayingAtariDeep2013}. 
  We discuss examples of possible metrics for $R$ and their decompositions into immediate rewards in Section~\ref{sec:exp_setup}. We now discuss how GPTRec can be trained to optimise any given effectiveness metric $R$.  
}

\begin{figure}[t]
    \centering
    \resizebox{\linewidth}{!}{
        \includegraphics{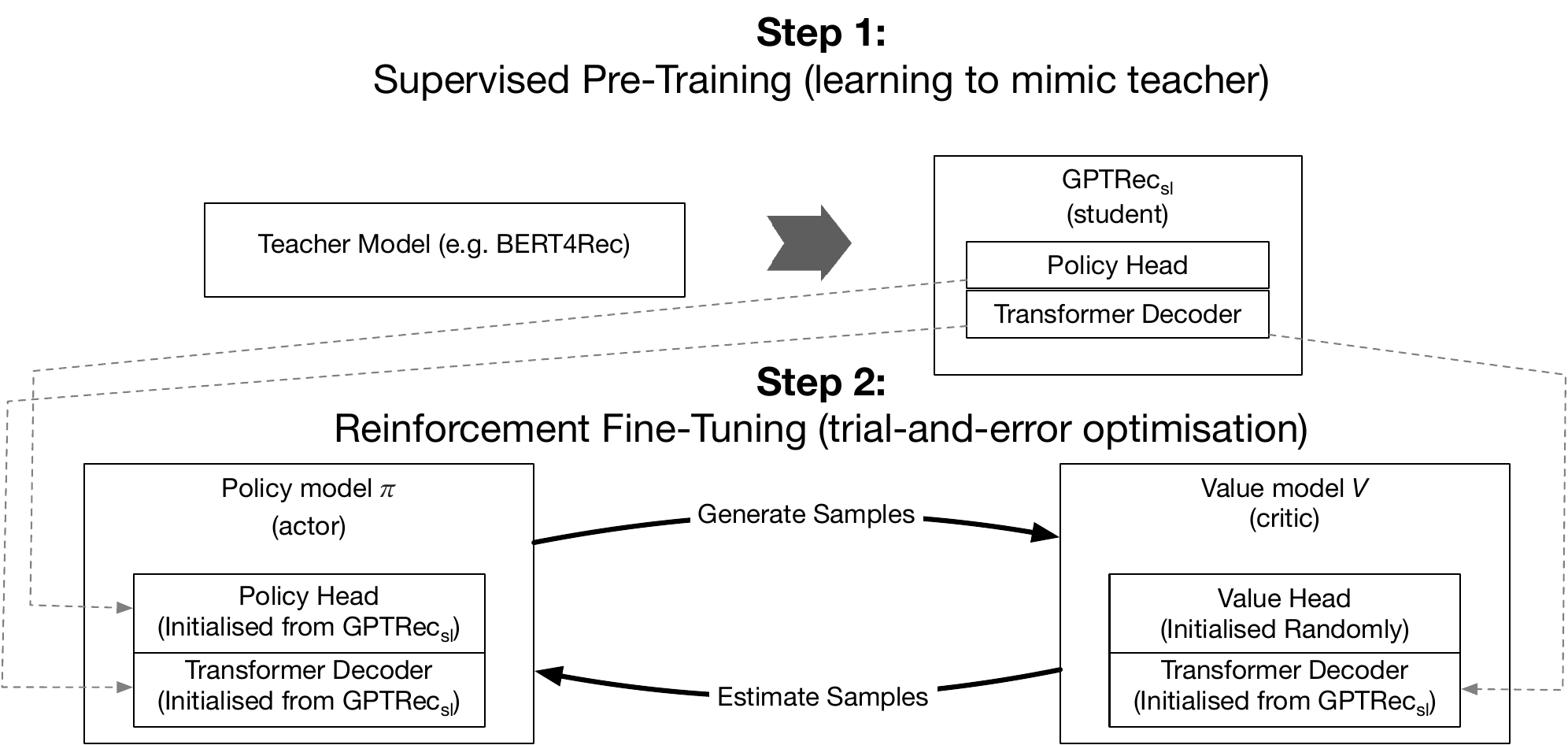}
    }
    \caption{GPTRec's Pre-Training/Fine-Tuning scheme. \sjj{
Pre-training (Step 1) takes the form of using a Top-K model like BERT4Rec (teacher) to pre-train a GPTRec model checkpoint (student). In Step 2 (Fine-tuning), the Policy model $\pi$ is the GTPRec model itself initialised by the student model checkpoint from Stage 1; the Value model is a Transformer Decoder-based model with a regression head. The Transformer Decoder layer of the Value model is initialised from the Transformer Decoder layer of the student model, and the regression head is initialised randomly.}}
    \label{fig:RLScheme}
\end{figure}

\slw{When GPTRec is used with the Next-K strategy, it outputs recommendations item-by-item. Consider a partially generated list of recommendations $g^{(i)}$ for the sequence $h$. 
\sdd{As} GPTRec is a probabilistic model, \sdd{it} estimates the conditional probability of an item appearing next in the recommendation list: 
\begin{align}
    \pi_\Theta(h, g^{(i)}, g_{i+1}) \approx  P(g_{i+1}|g^{(i)}; h)
\end{align}
where $\Theta$ is the set of learnable model parameters. 
\sj{Here, we use the symbol $\pi$ for the GPTRec model itself to highlight that this is used as a policy model in the Reinforcement Learning setup.} 
To understand which item \sdd{to} put to the position $i+1$, \sdd{the model obtains} the conditional probability distribution $\pi_\Theta$ and \sdd{selects} the item with the maximum probability:}
    \begin{align}
        g_{i+1} = \arg \max_{g_{i+1} \in I}\pi_\Theta(h, g^{(i)}, g_{i+1})
    \end{align}
Consider that we have a training set containing  "perfect" recommendations $\hat{g} = \{\hat{g}_1, \hat{g}_2, \hat{g}_3, .. \hat{g}_k\}$ for the sequences of historical interactions $h$  (the lists of recommendations, which maximise a given effectiveness measure $R(h, g)$ given $h$:
\begin{align}
    \hat{g} = \argmax_{g}R(h, g) \label{eq:perfect}
\end{align}
\sdm{Then, GPTRec can be optimised} using \sdm{the} standard Language Modelling (LM) loss (the same way, as  GPT~\cite{gpt2} is optimised given a large dataset of texts): 

\slw{
\begin{align}
    \mathcal{L}_{LM}(\Theta) = -\sum_{i\in 1.. k}\log \pi_\Theta(h, \hat{g}^{(i-1)}, \hat{g}_i) \label{eq:lmloss}
\end{align}
}
\slw{Unfortunately, the "perfect" list of recommendations, $\hat{g}$, is usually unknown. Indeed, given the complex structure of many possible quality metrics for $R$, in the general case, we have to score every possible recommendation list to find the perfect list $\hat{g}$ according to Equation~\eqref{eq:perfect}, which is not feasible due to the combinatorial size of the set of all possible recommendation lists. }

\slw{To solve this problem and inspired by the success of InstructGPT~\cite{ouyangTrainingLanguageModels2022} for solving a similar problem in the language processing domain, we propose a two-stage pre-training/fine-tuning approach.}
\looseness -1 \slw{
Figure~\ref{fig:RLScheme} illustrates the overall idea of the two-stage approach. As the figure shows, at the first stage, we use supervised pre-training to train the model to mimic the behaviour of a traditional Top-K recommendation model (a \emph{teacher} model), such as BERT4Rec~\cite{BERT4Rec}.  To do that, we first train the teacher using the training set; then, for each training sequence $h_j$, we generate a list of teacher recommendations $\Tilde{g}_j$; we then use $\Tilde{g}_j$ as the "perfect" recommendation $\hat{g}_j$ in \sdd{Equation~\eqref{eq:lmloss}} and optimise GPTRec using the LM loss. This process drives the model to generate recommendations \sdd{that} are as similar as possible to the ones generated by the teacher model. The problem is that this learning goal is not necessarily aligned with optimising the target metric $R$. Indeed, most traditional Top-K models are trained to optimise accuracy. At the same time, $R$ may include other components, such as diversity; \sw{therefore, a}  model pre-trained this way is likely to be sub-optimal with respect to $R$.}

\sasha{
    \sw{Hence}, in the second stage, as illustrated in Figure~\ref{fig:RLScheme}, we use a reinforcement learning-based approach to align GPTRec with the target metric $R$. 
    To achieve this, following the InstructGPT paper~\cite{ouyangTrainingLanguageModels2022}, we use the Proximal Policy Optimisation  (PPO) algorithm~\cite{schulmanProximalPolicyOptimization2017a}.
    \sjj{PPO is an} Actor-Critic type of approach~\cite[Ch. 15.1]{suttonReinforcementLearningIntroduction2018}, which jointly trains the main \emph{policy model $\pi$} ("Actor") and an auxiliary \emph{value model $V$} ("Critic"). Similar to other reinforcement learning methods, PPO is \sdd{a} trial-and-error approach, where the  Actor iteratively performs actions (generates recommendations) with the goal of beating the expectations of the Critic, and the Critic tries to predict the performance of the Actor in each case.}
    
In our case, we use GPTRec itself as the policy model $\pi$ and another GPTRec-based value model as the value model $V$. The difference between the \sj{policy} and \sj{value} models is that the \sj{policy} generates a probability distribution over all possible items in the catalogue, whereas the value model generates a single number: \sdd{a} predicted recommendation effectiveness according to metric $R$. 
\pageenlarge{3}
\subsection{Pre-Train/Fine Tune approach} \label{ssec:pre_train_fine_tune}

\begin{figure}
    \centering
    \resizebox{0.8\linewidth}{!}{
    \includegraphics{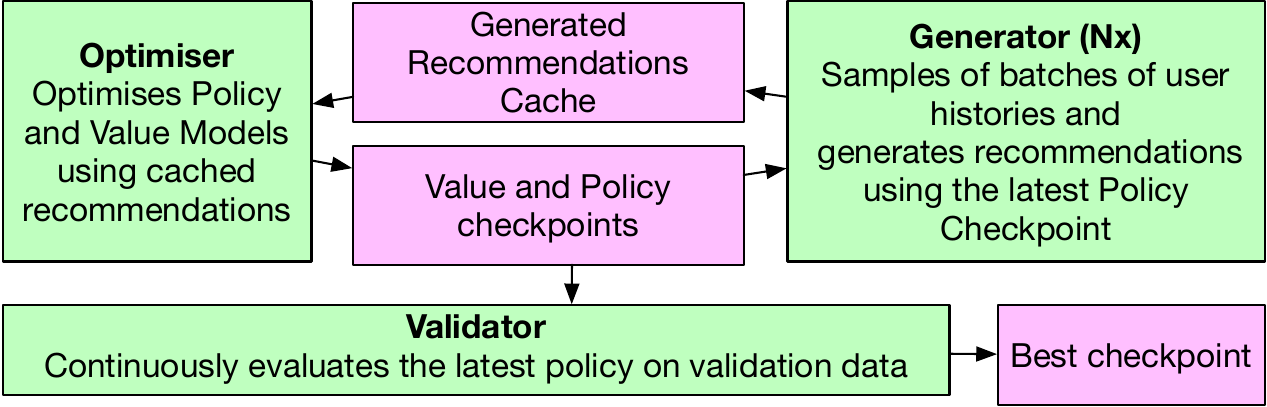}
    }
    \caption{\www{GPTRec fine-tuning processes diagram. Green boxes are processes; purple boxes are data.}}
    \label{fig:training_arch}
\end{figure}

Despite having different outputs, the input to both models has the same structure (illustrated in Figure~\ref{fig:sequence_structure}). Therefore, in both cases, we use the same GPT-2-based Transformer Decoder as the model backbone; however, in the value model, we use a simple feed-forward layer to predict a single number instead of a standard GPT-2 token prediction head. Following best practices~\cite{silverMasteringGameGo2016}, we initialise the policy model $\pi$ from the supervised checkpoint \sdd{trained in} the first stage ($GPTRec_{sl}$ in Figue~\ref{fig:training_arch}). We also initialise the backbone Transformer Decoder of the value model $V$ model from the backbone Transformer Decoder of the same pre-trained student model.  We then fine-tune both policy and value models using the PPO algorithm.  We provide a high-level overview of Reinforcement Learning in general and PPO in particular in Appendix~\ref{sec:rl} and refer to the original  \sdd{PPO} paper~\cite{schulmanProximalPolicyOptimization2017a} for more details of the approach.

\sasha{\subsection{Efficient Asynchronous Decomposition of Reinforcement Fine-Tuning} \label{ssec:rl_decompostion} }
\sasha{
    As we discussed in Section~\ref{ssec:pre_train_fine_tune}, \sdd{the PPO algorithm alternates between generating} recommendations using the current generation of the policy model and improving the policy and value models using gradient descent. Improving \sdd{the policy and the value} models using gradient descent can be efficiently parallelised using \sdd{ hardware acceleration on a GPU and a} deep learning framework (we use Tensorflow~\cite{tensorflowdevelopersTensorFlow2023})\footnote{\sdd{The code for this paper:  \href{https://github.com/asash/gptrec\_rl}{https://github.com/asash/gptrec\_rl}}}; however, generating recommendations with the Next-K strategy is an iterative process (recommendation for position $i+1$ can be only generated after \sdd{generating the} recommendation for position $i$). This iterative nature of generation limits the efficiency of using \sdd{the} GPU for generation. In initial experiments, we observed that directly alternating between generation and optimisation steps, \sdd{the} GPU becomes under-utilised, and model training becomes too slow to be feasible in practice (in our experiments, fine-tuning a single mode could take up to a week). 
    However, with PPO, we can make generation and optimisation asynchronous. Therefore, to efficiently utilise available \sdd{resources}, we \sdd{decompose} sample generation into a separate group of processes and perform the generation using the Next-K strategy on \sdd{the} CPU. We also use a separate validation process, which continuously evaluates the latest model checkpoint on a validation dataset. The validation \sdd{results} are used to monitor metrics during fine-tuning as well as in order to select the final model checkpoint. 
}
\begin{table*}[tb]
    \caption{Reinforcment learning optimisation objectives we use in versions of GPTRec. $y_i$ corresponds to the ground truth relevance for an item generated at the position $i$, $\text{CosDist}(g_i, g_j)$ is the cosine distance between the genres of items generated at the positions $i$ and $j$,  $\text{freq}(g_i)$ is the frequency in the training data of the item generated at the position $i$ and $\lambda$ is the secondary metric weitgh} \label{tb:rl_objectives}
    \vspace{-1\baselineskip}
    \begin{tabular}{c c}
          \toprule
          Model &  Learning Objective \\
          \midrule
          GPTRec-Reinforcement-NDCG &  $\sum_{i=1}^{K} \frac{y_i}{\log(i+1)}$ \\
          GPTRec-Reinforcement-Diversity &   $\sum_{i=1}^{K} \left[\frac{y_i}{\log(i+1)} + \lambda\cdot\frac{1}{K\cdot(K-1)}\sum_{j=1}^{i-1}\text{CosDist}(g_i, g_j)\right]$ \\
          GPTRec-Reinforcement-pCOUNT & $\sum_{i=1}^{K} \left[\frac{y_i}{\log(i+1)} - \lambda\cdot\text{freq}(g_i)\right]$ \\
          \bottomrule
    \end{tabular}

\end{table*}
\pageenlarge{3}
\sasha{
    Figure~\ref{fig:training_arch} summarises the processes involved in fine-tuning GPTRec.
    As can be seen from the figure, to train GPTRec, we use 3 processes:
        (1) \emph{\sdd{A Generator}} process that continuously samples batches of user histories from the training set and generates recommendations for these histories using the latest version of the policy model\sw{. After generating the recommendation, it saves the batch of histories and associated generated recommendations $\sdd{\langle h, g \rangle}$ pairs into the generated recommendations cache. The recommendation cache stores \sdd{the} last $M$ generated batches, where $M$ is a hyperparameter. Multiple Generators can work simultaneously; }
        (2) \emph{\sdd{An Optimiser}} that continuously samples batches of $\sdd{\langle h, g \rangle}$ pairs from the recommendations cache and performs optimisation steps for policy and value models using these \sdd{batches}. It then saves checkpoints of \sdd{the} Value and Policy models so the Generator and Validator processes can use them;
        (3) \emph{\sdd{A Validator}} that continuously takes the latest checkpoint and evaluates it using the validation dataset. 
    This scheme allows us to use both CPU and GPU system resources efficiently and reduces the training time of a single model from a few days to a few hours.}

\sw{This concludes the description of the proposed training scheme. We now to experimental evaluation of GPTRec trained with it.}

\section{Experimental Setup}\label{sec:exp_setup}

We aim to answer the following research questions: 
\www{
    \begin{enumerate}[font={\bfseries}, label={RQ\arabic*}, wide, labelwidth=!, labelindent=0pt]
        \item How effective is GPTRec trained as a supervised student compared to state-of-the-art baselines? \label{rq:student_effective}
        \item \sj{What is the effect of the teacher-student   pretraining scheme on the effectiveness of the Next-K strategy at different ranking cutoffs $K$?} \label{rq:importance_of_pretraining}
        \item \sj{What is the effect of RL-based tuning of GPTRec with Next-K generation when optimising for NDCG?} \label{rq:importance_of_rl}
        \item \sj{How does RL-based tuning affect beyond-accuracy metrics compared to greedy reranking?} \label{rq:baselines_comparison}
    \end{enumerate}
}

\subsection{Implementation}
 \sasha{We implement GPTRec using the GPT-2 architecture from  HuggingFace Transformers~\cite{wolfHuggingFaceTransformersStateoftheart2020} and the aprec\footnote{\href{https://github.com/asash/bert4rec\_repro}{https://github.com/asash/bert4rec\_repro}} framework from a recent replicability study~\cite{Bert4RecRepro}. We follow the experimental setup described in the same replicability study~\cite{Bert4RecRepro} and use the results of SASRec and BERT4Rec from that paper as baselines. \sw{Compared to the original GPTRec model, we use a slightly modified version of sequence structure, details of which we provide in in Appendix~\ref{ssec:sequence_structure}}.}
\begin{table}[tb]
    \caption{Salient characteristics of experimental datasets}\label{tb:dataset} 
    \resizebox{\linewidth}{!}{
    \www{
        \begin{tabular}{lrrrrrrr}
\toprule
\makecell[l]{Dataset} & \makecell[r]{Users} & \makecell[r]{Items} & \makecell[r]{Interactions} & \makecell[r]{Mean\\Length} & \makecell[r]{Median\\Length} & \makecell[r]{Sparsity} & \makecell[r]{Genres} \\
\midrule
MovieLens-1M & 6040 & 3416 & 999611 & 165.49 & 96 & 0.95 & 18 \\
Steam-2M & 201963 & 1000 & 2198260 & 10.88 & 7 & 0.99 & 318 \\
\bottomrule
\end{tabular}
}
    }
\end{table}

\begin{table*}[tb]
    \caption{\www{Evaluation results. The best results are highlighted in bold; 
            second best are underlined; * denotes significant difference with BERT4Rec ($pvalue < 0.05$, Bonferroni multi-test correction), arrows indicate the metric improvement direction}}\label{tb:results}
    \vspace{-1\baselineskip}
    \subfloat[MovieLens-1M]{
        \resizebox{\textwidth}{!}{
            \www{
                \begin{tabular}{lllllllll}
\toprule
Model Type  &  Model  &  \makecell[l]{Secondary\\Metric}  &  \makecell[l]{Secondary\\metric \\ weight ($\lambda$)}  &  Recall@1 	$\uparrow$  &  Recall@10 $\uparrow$  &  NDCG@10 $\uparrow$  &  \makecell[l]{Diversity \\ (ILD@10)}$\uparrow$   &  \makecell[l]{Popularity \\Bias \\ (nPCOUNT@10)} $\downarrow$ \\
\midrule
\multirow[t]{5}{*}{Baselines} & Popularity &   &   & 0.0056$^{<}$ & 0.0363$^{<}$ & 0.0178$^{<}$ & 0.3222$^{>}$ & 1.0000$^{>}$ \\
 & MF-BPR &   &   & 0.0113$^{<}$ & 0.0719$^{<}$ & 0.0366$^{<}$ & 0.2745 & 0.2867$^{<}$ \\
 & SASRec &   &   & 0.0464$^{<}$ & 0.2482$^{<}$ & 0.1306$^{<}$ & 0.2703$^{<}$ & \underline{0.2389$^{<}$} \\
 & GPTRec-Shifting &   &   & 0.0603 & 0.2647$^{<}$ & 0.1486$^{<}$ & 0.2749 & 0.2985$^{<}$ \\
 & BERT4Rec &   &   & 0.0608 & \textbf{0.2921} & 0.1617 & 0.2746 & 0.3222 \\
\cline{1-9}
Supervised & GPTRec-Supervised (MC Teacher) &   &   & 0.0791$^{>}$ & 0.2690$^{<}$ & 0.1638 & 0.2798$^{>}$ & 0.4470$^{>}$ \\
\cline{1-9}
NDCG Tuned & GPTRec-Reinforcement-NDCG &   &   & \textbf{0.0829$^{>}$} & \underline{0.2785} & \textbf{0.1682} & 0.2878$^{>}$ & 0.4205$^{>}$ \\
\cline{1-9}
\multirow[t]{2}{*}{Diversity tuned} & GPTRec-Reinforcement-Diversity-0.2 & ILD & 0.2 & 0.0798$^{>}$ & 0.2298$^{<}$ & 0.1499$^{<}$ & \underline{0.3621$^{>}$} & 0.5212$^{>}$ \\
 & GPTRec-Reinforcement-Diversity-1.0 & ILD & 1.0 & 0.0795$^{>}$ & 0.1748$^{<}$ & 0.1272$^{<}$ & \textbf{0.4349$^{>}$} & 0.5016$^{>}$ \\
\cline{1-9}
\multirow[t]{2}{*}{Tuned to decrease popularity bias} & GPTRec-Reinforcement-pCOUNT-3.0 & PCOUNT & 3.0 & \underline{0.0820$^{>}$} & 0.2733$^{<}$ & \underline{0.1669} & 0.2774 & 0.3936$^{>}$ \\
 & GPTRec-Reinforcement-pCOUNT-6.0 & PCOUNT & 6.0 & 0.0710 & 0.2154$^{<}$ & 0.1392$^{<}$ & 0.2648$^{<}$ & \textbf{0.1742$^{<}$} \\
\cline{1-9}
\bottomrule
\end{tabular}

            }
        }
    }
    \\ \vspace{-0.8\baselineskip}
    \subfloat[Steam-2M]{
        \resizebox{\textwidth}{!}{
            \www{
                \begin{tabular}{lllllllll}
\toprule
Model Type  &  Model  &  \makecell[l]{Secondary\\Metric}  &  \makecell[l]{Secondary\\metric \\ weight ($\lambda$)}  &  Recall@1 	$\uparrow$  &  Recall@10 $\uparrow$  &  NDCG@10 $\uparrow$  &  \makecell[l]{Diversity \\ (ILD@10)}$\uparrow$   &  \makecell[l]{Popularity \\Bias \\ (nPCOUNT@10)} $\downarrow$ \\
\midrule
\multirow[t]{5}{*}{Baselines} & Popularity &   &   & 0.0113$^{<}$ & 0.0725$^{<}$ & 0.0368$^{<}$ & 0.3282$^{>}$ & 1.0000$^{>}$ \\
 & MF-BPR &   &   & 0.0103$^{<}$ & 0.0570$^{<}$ & 0.0294$^{<}$ & 0.2923$^{<}$ & \underline{0.1093$^{<}$} \\
 & SASRec &   &   & 0.0197$^{<}$ & 0.1139$^{<}$ & 0.0588$^{<}$ & 0.3135 & 0.4826$^{<}$ \\
 & GPTRec-Shifting &   &   & 0.0329 & \underline{0.1719} & \underline{0.0912} & 0.3114$^{<}$ & 0.4301$^{<}$ \\
 & BERT4Rec &   &   & \underline{0.0331} & \textbf{0.1745} & \textbf{0.0923} & 0.3147 & 0.4973 \\
\cline{1-9}
\multirow[t]{2}{*}{Supervised} & GPTRec-Supervised (MC Teacher) &   &   & 0.0187$^{<}$ & 0.1060$^{<}$ & 0.0546$^{<}$ & 0.3210$^{>}$ & 0.8790$^{>}$ \\
 & GPTRec-Supervised (BERT4Rec Teacher) &   &   & 0.0326 & 0.1699 & 0.0908 & 0.3149 & 0.5041$^{>}$ \\
\cline{1-9}
NDCG Tuned & GPTRec-Reinforcement-NDCG &   &   & 0.0328 & 0.1709 & 0.0912 & 0.3148 & 0.5029$^{>}$ \\
\cline{1-9}
\multirow[t]{2}{*}{Diversity tuned} & GPTRec-Reinforcement-Diversity-1.0 & ILD & 1.0 & 0.0325 & 0.1530$^{<}$ & 0.0837$^{<}$ & \underline{0.3636$^{>}$} & 0.4825$^{<}$ \\
 & GPTRec-Reinforcement-Diversity-3.0 & ILD & 3.0 & 0.0303 & 0.1174$^{<}$ & 0.0699$^{<}$ & \textbf{0.4205$^{>}$} & 0.4744$^{<}$ \\
\cline{1-9}
\multirow[t]{2}{*}{Tuned to decrease popularity bias} & GPTRec-Reinforcement-pCOUNT-3.0 & PCOUNT & 3.0 & \textbf{0.0346} & 0.1606$^{<}$ & 0.0878$^{<}$ & 0.3141 & 0.3355$^{<}$ \\
 & GPTRec-Reinforcement-pCOUNT-6.0 & PCOUNT & 6.0 & 0.0245$^{<}$ & 0.0964$^{<}$ & 0.0563$^{<}$ & 0.3182$^{>}$ & \textbf{0.1049$^{<}$} \\
\cline{1-9}
\bottomrule
\end{tabular}

            }
        }
    }
\vspace{-1\baselineskip}
\end{table*}

\pageenlarge{3}
\subsection{Datasets}
\looseness -1 
 \sjj{As discussed in Section~\ref{sec:intro}, the scope of this work is limited to the datasets with a few thousand items in the catalogue. Indeed, there are solutions to allow Transformer models to scale to large catalogues, for instance, by using multiple tokens to represent each item ({\em sub-item representation})~\cite{GPTRec,petrovRecJPQTrainingLargeCatalogue2024} -- specifically,  original GPTRec paper~\cite{GPTRec} showed that GPTRec could be used with sub-item representations. Adding sub-item representations is a further experimental variable outside the scope of this paper.}

 Hence, we use the MovieLens-1M and Steam-2M datasets for our experiments, which both have $<$ 4000 items. While there are some known issues with the MovieLens-1M dataset~\cite{harperMovieLensDatasetsHistory2015}  (e.g.\ users don't rate movies in the dataset in the same order as they watch them), the dataset is used consistently across many papers, allowing comparison with published results. We use the version of the dataset from the SASRec~\cite{SASRec} repository and do not apply any additional preprocessing. The same version was used in the original GPTRec paper~\cite{GPTRec}  as well as many other recent publications~\cite{SASRec, BERT4Rec, Bert4RecRepro, petrovGSASRecReducingOverconfidence2023}.  We also use Steam-2M, a smaller version of the Steam~\cite{SASRec} dataset. In this version, we only use the 1000 most popular items.  Additionally, we de-duplicate the interactions in this dataset and remove users with less than 5 interactions. \sdm{Overall, Steam-2M has $2\times$ more interactions than Movielens-1M and $33\times$ more users, allowing us to test the model in a larger setting.} The salient characteristics of both experiments are listed in Table~\ref{tb:dataset}.

\subsection{Data Splitting}
We employ a leave-one-out strategy for model testing, holding out each user's latest interaction in the test set. In addition, for 128 randomly selected users, we hold out their second-to-last action in a separate validation set. This validation set is used for the early stopping mechanism and to control model quality during training, ensuring that our model generalises well to new data.

\sasha{Following common practice \sdd{in recommendation}~\cite{SASRec, BERT4Rec, Bert4RecRepro}, we use a relatively shallow \sdd{Transformer} model with three \sdd{Transformer} blocks. We set the maximum sequence length to \sasha{200} and truncate the sequence to the last \sasha{200} interactions if the user interacted with more items. We use 256-dimensional \sdd{item} embeddings in our experiments. At the supervised pre-training stage, we use a 200-epoch early stopping mechanism to ensure model convergence. At the fine-tuning stage, we tune the model for 64000 steps. We select the best model according to the target metric $R$ on validation data.}

\pageenlarge{3}
 \subsection{Effectiveness Metrics and Optimisation Goals} 
\sasha{We use NDCG@10 as the main accuracy metric of the models. We use PCOUNT~\cite{borgesMitigatingPopularityBias2021a} (\sdm{mean recommended item popularity}) as the popularity bias metric. To make PCOUNT independent of the dataset, we also use nPCOUNT --- a normalised version of the PCOUNT metric, in which we divide the PCOUNT metric by the PCOUNT metric of the popularity models. This way, the item popularity model (which has maximum possible popularity bias) always has an nPCOUNT value of 1, and the smaller values nPCOUNT of nPCOUNT correspond to the smaller popularity bias of the model (i.e., lower is better). Additionally, we use the Intra-List Distance (ILD) as the metric of the diversity of the recommended items~\cite{antikaciogluNewSystemWideDiversity2019a}. ILD is measured as a sum of pair-wise distances of the recommended items, and as the distance metric, we use the cosine distance between movie genre vectors for MovieLens-1M and game genres for Steam-2M. As the immediate reward $\mathrm{r}(h; g^{(i)})$ for NDCG@10 we use discounted relevance ($\frac{y_i}{\log(i+1)}$), where $y_i$ is the ground truth relevance or $i^{th}$ recommendation $g_i$); for PCOUNT as an immediate reward we use the negative item popularity and for diversity as an immediate reward we use the sum of pairwise diversities $\frac{1}{K\cdot(K-1)}\sum_{j \in 1..i-1}\text{CosDist}(g_i, g_j)$. For each immediate reward, the sum over all recommendation positions returns the value of the full metric. For PCOUNT, the immediate reward is negative, as we want to minimise the metric instead of maximising it. Table~\ref{tb:rl_objectives} also summaries reinforcement learning objectives we use in versions of GPTRec}

\pageenlarge{3}
\subsection{Baselines} As the baseline models, we use two simple models: Item Popularity and BPR~\cite{BPR} (we use a version of BPR from the LightFM library). Additionally, we use \sj{several} \sdd{Transformer-based} models: (1) SASRec~\cite{SASRec}, the most cited transformer-based recommendation model.   SASRec is architecturally very similar to GPTRec (it is also Transformer Decoder-based) but uses the Top-K recommendation strategy. In contrast with GPTRec, it is trained using sequence-to-sequence shifting objectives. The loss function is \sjj{Binary Cross-Entropy};%
    (2) BERT4Rec is a popular Transformer Encoder-based baseline, which exhibits state-of-the-art effectiveness \sjj{when fully converged~\cite{Bert4RecRepro}.} \sjj{It uses the Softmax Cross-Entropy as the loss function};
    (3) GPTRec-Shifting is \sw{a GPTRec-based baseline, that has been trained using a traditional training scheme rather than the proposed 2-stage scheme}. Similar to SASRec, it is trained using the sequence shifting objective, but similarly to BERT4Rec, it uses the Softmax Cross-Entropy the loss function instead of the Binary Cross-Entropy. Indeed, \sjj{Softmax Cross-Entropy has been shown} to be more effective for sequential recommendation~\cite{petrovGSASRecReducingOverconfidence2023}. 
    Similar to GPTRec, GPTRec-Shifting uses GPT-2 architecture as the backbone. \sjj{The motivation for us to include GPTRec-Shifting as a baseline is to decouple the effect of model architecture change from the training scheme change when comparing GPTRec with the traditional models, such as SASRec and BERT4Rec.}

\vspace{-0.5\baselineskip}
\subsection{Teacher Models for Supervised Learning} 
\looseness -1 \sdd{For the first stage of the training process, we use two teacher models. First, we use a simple Markov Chain-based (MC) model, which calculates the empirical probability of item $a$ appearing in the sequence after item $b$ (immediately or after several steps). It then computes the final score of the weighted sum of empirical probabilities: $s(g_{i}) = \sum_{j \in 1..n} \beta^j \log p(g_i|h_{n-j+1}) $, where $\beta$ is a decay hyperparameter, \sdd{which we set to 0.6 (we find this to work the best experimentally)}. This teacher model is very simple, and training it only takes a few seconds on both datasets. Second, for the Steam-2M dataset, where using Markov Chain as a teacher resulted in weaker performance viz. the baselines, we also use BERT4Rec as a teacher.}%

\begin{figure*}
    \centering
    \subfloat[MovieLens-1M]{
        \resizebox{0.5\linewidth}{!}{
            \includegraphics{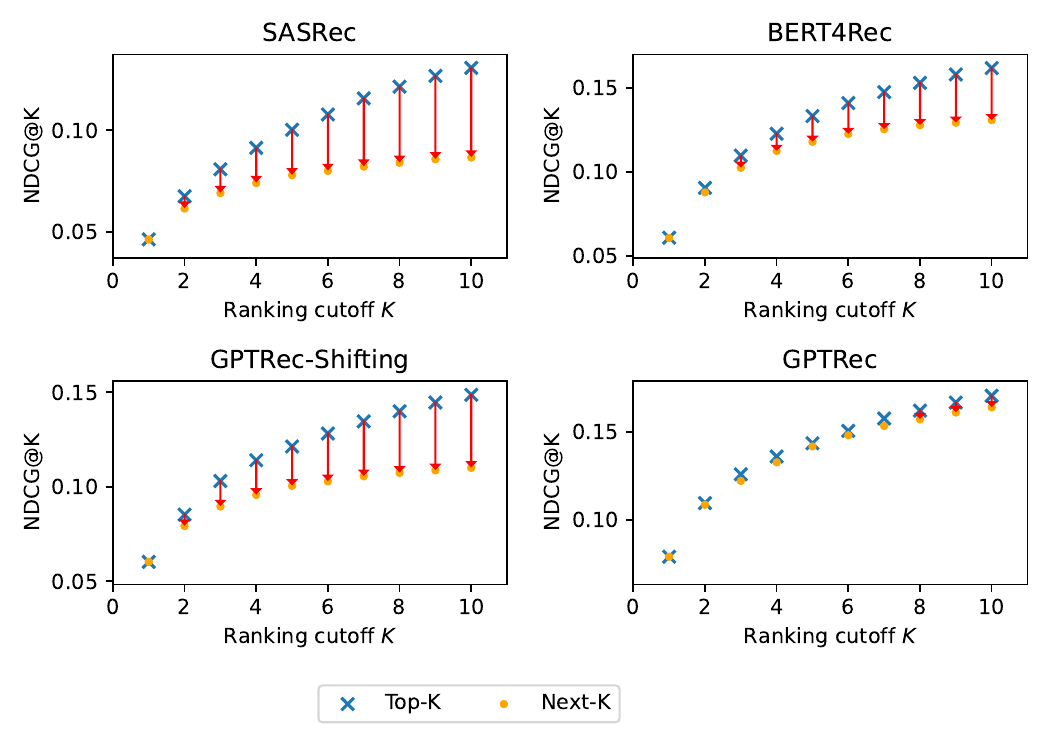}
        }
    } 
    \subfloat[Steam-2M]{
    \resizebox{0.5\linewidth}{!}{
            \includegraphics{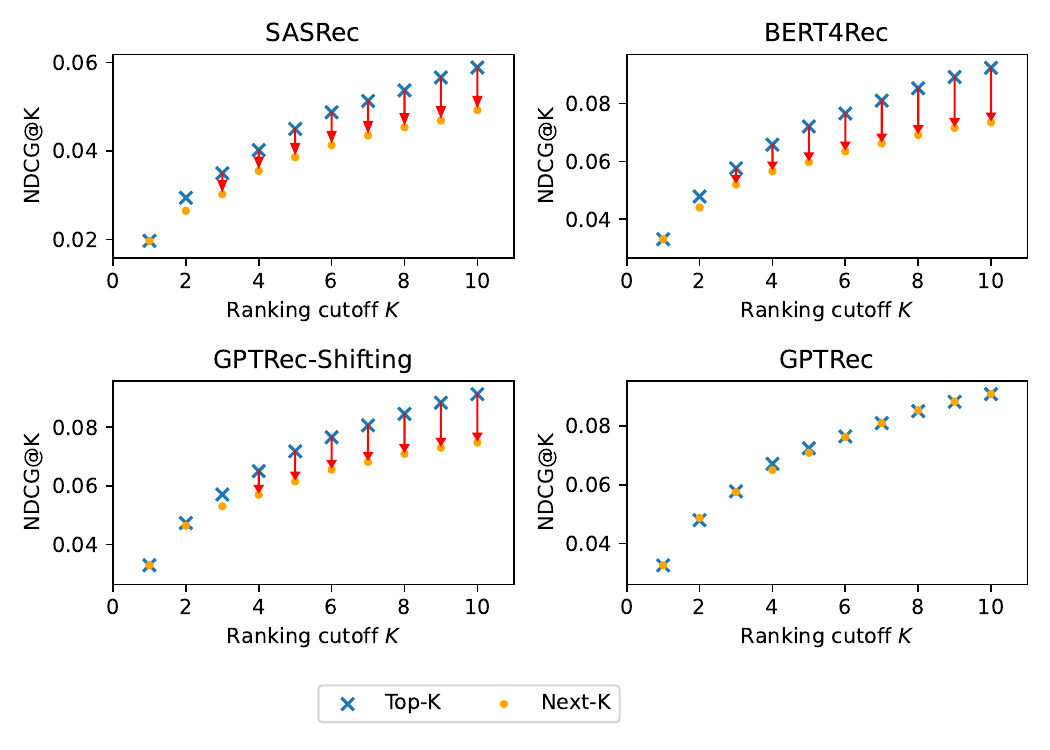}
        }
    } \\
    \caption{Models' NDCG@K with Top-K and Next-K recommendation strategies when varying ranking cutoff $K$. \sjj{Red arrows demonstrate the effectiveness gap between the Top-K and the Next-K inference strategies at a given ranking cutoff $K$.}}
    \label{fig:top_k_vs_next_k}
    \vspace{-1\baselineskip}
\end{figure*}

\vspace{-0.6\baselineskip}
\section{Results}\label{sec:results}
\vspace{-0.3\baselineskip}
\pageenlarge{3}
\subsection{\ref{rq:student_effective} Supervised-Only training} 
\looseness -1 \sasha{We first analyse the effectiveness of GPTRec when it is trained only using the first stage of the training process, without reinforcement learning.}
\sasha{
    Table~\ref{tb:results} contains the results of this comparison (see model types "Supervised" and "Baselines") for (a) MovieLens-1M and (b) Steam-2M.  As we can see from Table~\ref{tb:results}(a), for MovieLens-1M, the model with this simple teacher performs surprisingly well, achieving similar accuracy metrics to BERT4Rec (NDCG@10 0.1638 for GPTRec, 0.1617 for BERT4Rec, \sdd{difference} not significant). \sdd{Interestingly, on this dataset \sdd{GPTRec} also achieves much higher Recall@1 compared to BERT4Rec} (0.0791, +30.1\%). We speculate that the improvement is due to a better alignment of GPTRec's training task with the next item prediction task; see~\cite{petrovGSASRecReducingOverconfidence2023} for similar findings. On the other hand, on the Steam-2M dataset (Table~\ref{tb:results}(b)), GPTRec with the Markov \sdd{Chain} teacher has a lower effectiveness than BERT4Rec (NDCG@10 0.0546 vs.\  0.0923, significant); however, it is still comparable with SASRec, which has a slightly higher NDCG@10 of 0.0588. 
    Due to the weaker effectiveness of GPTRec compared to BERT4Rec on the Steam-2M dataset, we additionally train GPTRec using BERT4Rec as a teacher. In this case,  GPTRec achieves a similar NDCG@10 to BERT4Rec (1.6\% difference, not significant). \sjj{Hence, for steam dataset in all following experiments we use the version of GPTRec pre-trained with the BERT4Rec teacher.}
}

\sasha{
    Overall, in answer to~\ref{rq:student_effective}, we say that \sw{GPTRec pre-trained as supervised student}  can achieve similar accuracy (measured by NDCG@10) to the state-of-the-art BERT4Rec model and outperforms other baselines. Importantly, it achieves good effectiveness using the \emph{generative} Next-K strategy, meaning that the models from the first training stage can serve as a good starting point for the fine-tuning at the second stage.}
    
\pageenlarge{3}
\vspace{-0.8\baselineskip}
\subsection{\ref{rq:importance_of_pretraining} Effect of Teacher-Student pertaining on the Next-K recommendation.} \label{ssec:results:teacher_student_pretrainig}
\sj{
    \sw{In principle, } the Next-K strategy \sjj{can be used with any} sequential recommendation model, by iteratively adding a generated item to the end of the input sequence. However, we hypothesise the model needs to be trained specifically for Next-K generation -- i.e., using a Top-K model checkpoint leads to effectiveness degradation at large cutoffs $K$ for Next-K generation. To empirically validate the hypothesis, we compare the NDCG@$K$ metric at different cutoffs $K$ when using the Top-K and the Next-K \sjj{inference}. 
    We analyse the three baseline Transformer-based models (SASRec, BERT4Rec and GPTRec-Shifting) trained using their classical training objectives and GPTRec trained with the teacher-student scheme. We compare these baseline models using both Top-K and Next-K generation inference strategies.
    \sjj{For all four models, we compare the same model checkpoint under different inference strategies; the only change is switching from Top-K inference to Next-K inference.}}

\sj{
    \looseness -1 Figure~\ref{fig:top_k_vs_next_k} shows the results of this analysis on both experimental datasets. As we can see from the figure, in all cases at cutoff $K=1$, both Top-K and Next-K strategies always achieve the same NDCG@1. Indeed, when we are only interested in one item, both strategies select the item with the maximum score according to the model, and hence both strategies are equal in case $K=1$.  However, at deeper ranking cutoffs, we observe quality effectiveness degradation when using the Next-K strategy with the baseline models. For example, with $K=10$, SASRec shows 33.7\% lower NDCG@10 when using the Next-K strategy compared to the Top-K. 
    In contrast, for GPTRec pre-trained with the teacher-student, the difference between Top-K and Next-K is usually small and sometimes positive. For example, when using ranking cutoff $K=8$, GPTRec with Next-K strategy achieves a slightly higher NDCG@8 (+0.35\%) on the Steam-2M dataset while achieving slightly lower NDCG@8 on MovieLens-1M dataset (-2.99\%). 
}
\pageenlarge{3}
\sj{
    \sjj{When comparing the plots of the GPTRec with GPTRec-Shifting (which uses the same backbone architecture but classic sequence shifting training), we observe markedly different trends.    
    Indeed, on both datasets, similarly to other baselines, GPTRec-Shifting exhibits effectiveness degradations when switching from the Top-K to the Next-K inference strategy. However, for GPTRec, Top-K and Next-K exhibit widely similar effectiveness across $K$ values. For example, we observe a -16.14\% NDCG@8 drop when switching from Top-K to Next-K at ranking cutoff $K=8$ on Steam-2M, which has a similar magnitude to the other baseline models (-15.59\% for SASRec, -18.93\% for BERT4Rec). In contrast, for the same dataset and at the same cutoff $K=8$, GPTRec exhibits a +0.35\% improvement. This shows that the absence of effectiveness degradation at deeper cutoffs $K$ when using the Next-K strategy compared to Top-K in GPTRec is caused by the training scheme from sequence shifting to the teacher-student pre-training and not by the architecture (GPTRec \& GPTRec-Shifting share the backbone architecture but perform differently).}}

\looseness -1 \sj{Overall, for \ref{rq:importance_of_pretraining}, we conclude that a teacher-student training scheme is necessary to achieve the same effectiveness, using the Next-K strategy, as it is achievable with the Top-K strategy. Hence Next-K is a viable strategy for recommendation, but requires an appropriate training scheme. In the next sections, we empirically demonstrate its benefits compared Top-K for accuracy and beyond-accuracy metrics as a result of applying reinforcement learning.}

\subsection{\ref{rq:importance_of_rl}. Reinforcement Learning for Accuracy}
\sasha{
    \looseness -1 We now analyse the effectiveness of the \sdd{full} 2-stage training scheme when our goal is to optimise accuracy, i.e., using NDCG@10 as the effectiveness measure $R$. For MovieLens-1M, we start from the checkpoint which was trained using the Markov Chain teacher, and for Steam-2M, we use the checkpoint trained using the BERT4Rec teacher; both these checkpoints achieve competitive results to BERT4Rec even without fine-tuning. 
    Table~\ref{tb:results} summarises the results (see "NDCG-Tuned" Model Type). As we can see from the table, fine-tuning increases NDCG@10 in both cases, but this improvement is very small (and not statistically significant). For example, on MovieLens-1M, NDCG@10 improved from 0.1638 to 0.1682 (+2.6\%, not significant), and on Steam-2M, it improved from 0.0908 to 0.0912 (+0.4\%, not significant). In both cases, the result is on par with BERT4Rec (statistically indistinguishable). 
    In summary, for~\ref{rq:importance_of_rl}, we observe a small positive effect of reinforcement learning tuning for NDCG, but this effect is small and not statistically significant. In practice, it is enough to use only first-stage training if the goal is to optimise the model for accuracy only (but not for more complex goals; see the next section).}

\begin{figure}[t!]
\subfloat[MovieLens-1M]{
    \includegraphics[width=0.49\linewidth]{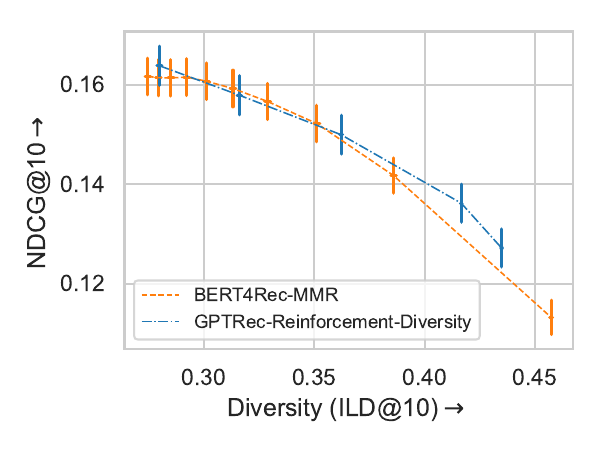}
}
\subfloat[Steam-2M]{
    \includegraphics[width=0.49\linewidth]{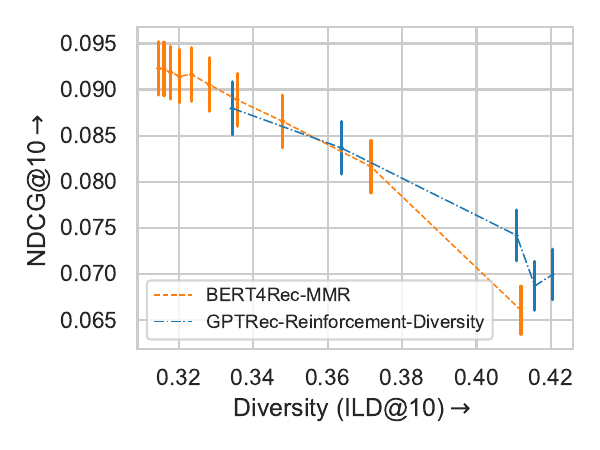}
}
\caption{\www{Accuracy (NDCG@10) / Diversity(ILD@10) tradeoff. Arrows represent the direction of metric improvement, and horizontal and vertical lines represent standard errors.}} \label{figl:ild_tradeoff}
\end{figure}

\subsection{\ref{rq:baselines_comparison}. Reinforcement Learning for Beyond-Accuracy Measures}

\begin{figure}[t]
    \subfloat[MovieLens-1M]{
        \includegraphics[width=0.45\linewidth]{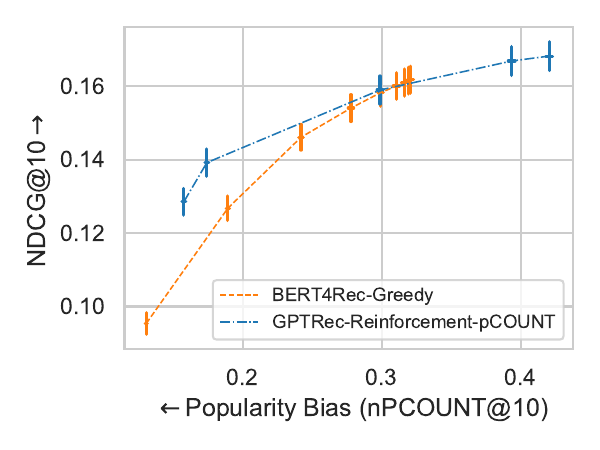}
    }
    \subfloat[Steam-2M]{
        \includegraphics[width=0.45\linewidth]{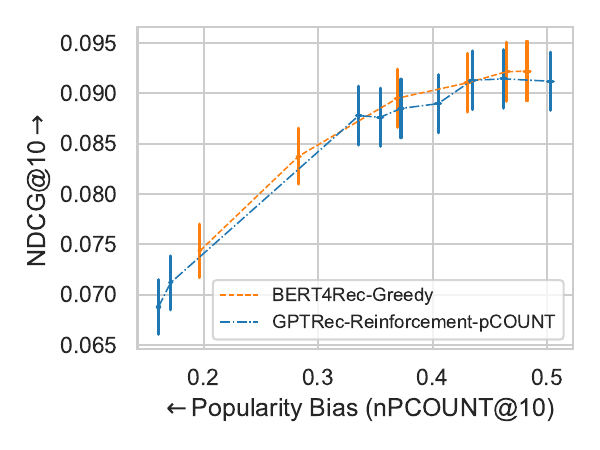}
    }
    \caption{\www{Accuracy (NDCG@10) / Popularity Bias (nPCOUNT@10) tradeoff.  Arrows represent the direction of metric improvement, and horizontal and vertical lines represent standard errors.}} \label{fig:popbias_tradeoff}
\end{figure}

\pageenlarge{3}
\looseness -1 In our last research question, we analyse the effect of the 2nd-stage tuning when the effectiveness metric $R$ includes additional components. In our experiments, we use the compositional effectiveness metric $R = NDCG + \lambda \cdot R_{Secondary}$, where $R_{Secondary}$ measures diversity (using ILD) or popularity bias (using negative PCOUNT). In our experiments, we vary $\lambda$ between 0 and 3 for ILD and between 0 and 6 for PCOUNT.  As the baseline for secondary metric optimisation, we use greedy reranking techniques over \sdd{the} BERT4Rec results: Maximal Marginal Relevance (MMR)~\cite{carbonellUseMMRDiversitybased1998b} as the diversity baseline,  and greedy re-ranking~\cite{kayaAccurateDiverseRecommendations2018} for decreasing popularity bias. In both cases, we control the tradeoff between primary and secondary metrics \sdd{in the baselenies} by varying the weight of the secondary metric during the re-ranking stage.
\looseness -1 \sasha{Figures~\ref{figl:ild_tradeoff} and~\ref{fig:popbias_tradeoff} illustrate the results of our \sdd{experiments}. As can be seen from the figures, the results achieved by GPTRec are not worse compared to MMR for diversity and compared to greedy re-ranking for PCOUNT. \sdm{Indeed, in all cases, when placing most optimisation emphasis on NDCG, the results are not distinguishable between BERT4Rec and GPTRec (we already discussed that in~\ref{rq:importance_of_rl})}. However, when we increase the \sdd{importance} of \sdd{the} secondary metrics, GPTRec outperforms greedy techniques over BERT4Rec in 3 out of 4 cases, with the only exception being tuning for decreasing popularity bias on Steam-2M. For example, when GPTRec is tuned to decrease popularity bias on MovieLens-1M, it achieves NDCG@10 of 0.139 and nPCOUNT of 0.1742. At the same time, BERT4Rec can only achieve lower NDCG@10 of 0.1267 (-8.8 \%, significant), with higher popularity bias (nPCOUNT 0.1893, +8.6\%, significant)}.

\pageenlarge{3}
We also provide the quantitative results of tuning the models \sdm{for beyond-accuracy metrics} in Table~\ref{tb:results}. As can be seen from the table, when the model is tuned for a specific secondary metric, the model is able to achieve good results according to this metric while retaining reasonably high NDCG@10. For example, when we tune GPTRec for diversity on the Steam-2M dataset (Table~\ref{tb:results}(b)), with the weight of secondary metric $\lambda=6.0$, we obtain the best ILD (0.4205, +33\% over BERT4Rec, significant) while still retaining NDCG@10 comparable to SASRec.
    
Overall, in answer to~\ref{rq:baselines_comparison}, we conclude that fine-tuning allows us to optimise GPTRec \sdm{with the Next-K generation strategy} for complex recommendation goals, and the results are better (in $\frac{3}{4}$ cases) or \sjj{statistically} indistinguishable ($\frac{1}{4}$ case) compared to greedy reranking of the BERT4Rec results.

\section{Conclusions} \label{sec:conclusion}
\looseness -1 \sdm{\sw{In this paper, we proposed a 2-stage pre-training/fine-tuning approach to align generative sequential models with beyond-arccuracy goals,  where at the pre-training stage, a generative model learns to mimic the behaviour of traditional Top-K recommenders, and during the fine-tuning stage, it can be optimised for any metric using reinforcement learning. We applied this approach to the recently proposed GPTRec model, which uses the Next-K strategy for recommendations. 
We demonstrated that the first stage (supervised pre-training) is enough to achieve NDCG@10 comparable to the state-of-the-art BERT4Rec model;} however, fine-tuning can improve secondary objectives, such as diversity and popularity bias. In 3 out of 4 experiments, GPTRec fine-tuned for complex metrics, including diversity and popularity bias, outperformed the greedy re-ranking over BERT4Rec results. In 1 out of 4, the results were indistinguishable. For example, a fine-tuned version of GPTRec allowed us to simultaneously achieve 8.8\% better NDCG and 8.6\% lower popularity bias compared to greedy reranking over BERT4Rec results.}

\looseness -1 \sjj{The methodology proposed in this allows direct optimisation of GPTRec (or, indeed, any other Transformer Decoder model) for any effectiveness measure without requiring the measure to be differentiable or computable at a single-item level. We demonstrated the applicability of this methodology using such beyond-accuracy objective targets as increasing diversity and decreasing popularity bias; however, the methodology can be used with other metrics, including variations of fairness measures, measures that balance the needs of various stakeholders of recommender systems and so on. } 

\sjj{There are a number of possible future research directions that can build upon this work: Firstly, the scope of applicability of the methodology can be extended to larger datasets by applying sub-item representations instead of using atomic item IDs; Secondly, the quality of the generated rankings could possibly be further be improved by enhancing the model architecture; Thirdly, the recommendation goal can be used as the model input, allowing a single model to generate different kinds of recommendations and providing the user control over the desired output. Overall, exploring these future research directions could further refine the effectiveness, efficiency and applicability of recommendation systems in many scenarios.}
\FloatBarrier
\bibliographystyle{ACM-Reference-Format}
\bibliography{references}


\begin{thebibliography}{36}


\ifx \showCODEN    \undefined \def \showCODEN     #1{\unskip}     \fi
\ifx \showDOI      \undefined \def \showDOI       #1{#1}\fi
\ifx \showISBNx    \undefined \def \showISBNx     #1{\unskip}     \fi
\ifx \showISBNxiii \undefined \def \showISBNxiii  #1{\unskip}     \fi
\ifx \showISSN     \undefined \def \showISSN      #1{\unskip}     \fi
\ifx \showLCCN     \undefined \def \showLCCN      #1{\unskip}     \fi
\ifx \shownote     \undefined \def \shownote      #1{#1}          \fi
\ifx \showarticletitle \undefined \def \showarticletitle #1{#1}   \fi
\ifx \showURL      \undefined \def \showURL       {\relax}        \fi
\providecommand\bibfield[2]{#2}
\providecommand\bibinfo[2]{#2}
\providecommand\natexlab[1]{#1}
\providecommand\showeprint[2][]{arXiv:#2}

\bibitem[Afsar et~al\mbox{.}(2023)]%
        {afsarReinforcementLearningBased2023}
\bibfield{author}{\bibinfo{person}{M.~Mehdi Afsar}, \bibinfo{person}{Trafford
  Crump}, {and} \bibinfo{person}{Behrouz Far}.}
  \bibinfo{year}{2023}\natexlab{}.
\newblock \showarticletitle{Reinforcement {{Learning}} Based {{Recommender
  Systems}}: {{A Survey}}}.
\newblock \bibinfo{journal}{\emph{Comput. Surveys}} \bibinfo{volume}{55},
  \bibinfo{number}{7} (\bibinfo{year}{2023}), \bibinfo{pages}{1--38}.
\newblock


\bibitem[Antikacioglu et~al\mbox{.}(2019)]%
        {antikaciogluNewSystemWideDiversity2019a}
\bibfield{author}{\bibinfo{person}{Arda Antikacioglu}, \bibinfo{person}{Tanvi
  Bajpai}, {and} \bibinfo{person}{R. Ravi}.} \bibinfo{year}{2019}\natexlab{}.
\newblock \showarticletitle{A {{New System-Wide Diversity Measure}} for
  {{Recommendations}} with {{Efficient Algorithms}}}.
\newblock \bibinfo{journal}{\emph{SIAM Journal on Mathematics of Data Science}}
  \bibinfo{volume}{1}, \bibinfo{number}{4} (\bibinfo{year}{2019}),
  \bibinfo{pages}{759--779}.
\newblock


\bibitem[Bai et~al\mbox{.}(2019)]%
        {baiModelbasedReinforcementLearning2019}
\bibfield{author}{\bibinfo{person}{Xueying Bai}, \bibinfo{person}{Jian Guan},
  {and} \bibinfo{person}{Hongning Wang}.} \bibinfo{year}{2019}\natexlab{}.
\newblock \showarticletitle{A Model-Based Reinforcement Learning with
  Adversarial Training for Online Recommendation}. In
  \bibinfo{booktitle}{\emph{Proc. {{NeurIPS}}}}. \bibinfo{pages}{10735--10746}.
\newblock


\bibitem[Borges and Stefanidis(2021)]%
        {borgesMitigatingPopularityBias2021a}
\bibfield{author}{\bibinfo{person}{Rodrigo Borges} {and}
  \bibinfo{person}{Kostas Stefanidis}.} \bibinfo{year}{2021}\natexlab{}.
\newblock \showarticletitle{On Mitigating Popularity Bias in Recommendations
  via Variational Autoencoders}. In \bibinfo{booktitle}{\emph{Proc. {{SAC}}}}.
  \bibinfo{pages}{1383--1389}.
\newblock


\bibitem[Carbonell and Goldstein(1998)]%
        {carbonellUseMMRDiversitybased1998b}
\bibfield{author}{\bibinfo{person}{Jaime Carbonell} {and} \bibinfo{person}{Jade
  Goldstein}.} \bibinfo{year}{1998}\natexlab{}.
\newblock \showarticletitle{The Use of {{MMR}}, Diversity-Based Reranking for
  Reordering Documents and Producing Summaries}. In
  \bibinfo{booktitle}{\emph{Proc. {{SIGIR}}}}. \bibinfo{pages}{335--336}.
\newblock


\bibitem[Deffayet et~al\mbox{.}(2023)]%
        {deffayetGenerativeSlateRecommendation2023}
\bibfield{author}{\bibinfo{person}{Romain Deffayet}, \bibinfo{person}{Thibaut
  Thonet}, \bibinfo{person}{Jean-Michel Renders}, {and}
  \bibinfo{person}{Maarten De~Rijke}.} \bibinfo{year}{2023}\natexlab{}.
\newblock \showarticletitle{Generative {{Slate Recommendation}} with
  {{Reinforcement Learning}}}. In \bibinfo{booktitle}{\emph{Proc. {{WSDM}}}}.
  \bibinfo{pages}{580--588}.
\newblock


\bibitem[Harper and Konstan(2015)]%
        {harperMovieLensDatasetsHistory2015}
\bibfield{author}{\bibinfo{person}{F.~Maxwell Harper} {and}
  \bibinfo{person}{Joseph~A. Konstan}.} \bibinfo{year}{2015}\natexlab{}.
\newblock \showarticletitle{The {{MovieLens Datasets}}: {{History}} and
  {{Context}}}.
\newblock \bibinfo{journal}{\emph{ACM Transactions on Interactive Intelligent
  Systems (TiiS)}} \bibinfo{volume}{5}, \bibinfo{number}{4}
  (\bibinfo{date}{Dec.} \bibinfo{year}{2015}), \bibinfo{pages}{19:1--19:19}.
\newblock
\showISSN{2160-6455}


\bibitem[Kang and McAuley(2018)]%
        {SASRec}
\bibfield{author}{\bibinfo{person}{Wang-Cheng Kang} {and}
  \bibinfo{person}{Julian McAuley}.} \bibinfo{year}{2018}\natexlab{}.
\newblock \showarticletitle{Self-{{Attentive Sequential Recommendation}}}. In
  \bibinfo{booktitle}{\emph{Proc. {{ICDM}}}}. \bibinfo{pages}{197--206}.
\newblock
\showISSN{2374-8486}


\bibitem[Kaya(2018)]%
        {kayaAccurateDiverseRecommendations2018}
\bibfield{author}{\bibinfo{person}{Mesut Kaya}.}
  \bibinfo{year}{2018}\natexlab{}.
\newblock \showarticletitle{Accurate and {{Diverse Recommendations Using
  Item-Based SubProfiles}}}. In \bibinfo{booktitle}{\emph{Proc. {{FLAIRS}}}}.
\newblock


\bibitem[Kotkov et~al\mbox{.}(2020)]%
        {kotkovHowDoesSerendipity2020}
\bibfield{author}{\bibinfo{person}{Denis Kotkov}, \bibinfo{person}{Jari
  Veijalainen}, {and} \bibinfo{person}{Shuaiqiang Wang}.}
  \bibinfo{year}{2020}\natexlab{}.
\newblock \showarticletitle{How Does Serendipity Affect Diversity in
  Recommender Systems? {{A}} Serendipity-Oriented Greedy Algorithm}.
\newblock \bibinfo{journal}{\emph{Computing}} \bibinfo{volume}{102},
  \bibinfo{number}{2} (\bibinfo{year}{2020}), \bibinfo{pages}{393--411}.
\newblock


\bibitem[Kunaver and Po{\v{z}}rl(2017)]%
        {kunaver2017diversity}
\bibfield{author}{\bibinfo{person}{Matev{\v{z}} Kunaver} {and}
  \bibinfo{person}{Toma{\v{z}} Po{\v{z}}rl}.} \bibinfo{year}{2017}\natexlab{}.
\newblock \showarticletitle{Diversity in recommender systems--A survey}.
\newblock \bibinfo{journal}{\emph{Knowledge-based systems}}
  \bibinfo{volume}{123} (\bibinfo{year}{2017}), \bibinfo{pages}{154--162}.
\newblock


\bibitem[Mnih et~al\mbox{.}(2013)]%
        {mnihPlayingAtariDeep2013}
\bibfield{author}{\bibinfo{person}{Volodymyr Mnih}, \bibinfo{person}{Koray
  Kavukcuoglu}, \bibinfo{person}{David Silver}, \bibinfo{person}{Alex Graves},
  \bibinfo{person}{Ioannis Antonoglou}, \bibinfo{person}{Daan Wierstra}, {and}
  \bibinfo{person}{Martin Riedmiller}.} \bibinfo{year}{2013}\natexlab{}.
\newblock \bibinfo{title}{Playing {{Atari}} with {{Deep Reinforcement
  Learning}}}.
\newblock
\newblock
\showeprint[arxiv]{1312.5602}~[cs]


\bibitem[Ouyang et~al\mbox{.}(2022)]%
        {ouyangTrainingLanguageModels2022}
\bibfield{author}{\bibinfo{person}{Long Ouyang}, \bibinfo{person}{Jeff Wu},
  \bibinfo{person}{Xu Jiang}, \bibinfo{person}{Diogo Almeida},
  \bibinfo{person}{Carroll~L. Wainwright}, \bibinfo{person}{Pamela Mishkin},
  \bibinfo{person}{Chong Zhang}, \bibinfo{person}{Sandhini Agarwal},
  \bibinfo{person}{Katarina Slama}, \bibinfo{person}{Alex Ray},
  \bibinfo{person}{John Schulman}, \bibinfo{person}{Jacob Hilton},
  \bibinfo{person}{Fraser Kelton}, \bibinfo{person}{Luke Miller},
  \bibinfo{person}{Maddie Simens}, \bibinfo{person}{Amanda Askell},
  \bibinfo{person}{Peter Welinder}, \bibinfo{person}{Paul Christiano},
  \bibinfo{person}{Jan Leike}, {and} \bibinfo{person}{Ryan Lowe}.}
  \bibinfo{year}{2022}\natexlab{}.
\newblock \bibinfo{title}{Training Language Models to Follow Instructions with
  Human Feedback}.
\newblock
\newblock
\showeprint[arxiv]{2203.02155}~[cs]


\bibitem[Petrov and Macdonald(2024)]%
        {petrovRecJPQTrainingLargeCatalogue2024}
\bibfield{author}{\bibinfo{person}{Aleksandr Petrov} {and}
  \bibinfo{person}{Craig Macdonald}.} \bibinfo{year}{2024}\natexlab{}.
\newblock \showarticletitle{{{RecJPQ}}: {{Training Large-Catalogue Sequential
  Recommenders}}}. In \bibinfo{booktitle}{\emph{Proc. {{WSDM}}}}.
\newblock


\bibitem[Petrov and Macdonald(2022a)]%
        {PetrovRSS22}
\bibfield{author}{\bibinfo{person}{Aleksandr~V. Petrov} {and}
  \bibinfo{person}{Craig Macdonald}.} \bibinfo{year}{2022}\natexlab{a}.
\newblock \showarticletitle{Effective and {{Efficient Training}} for
  {{Sequential Recommendation}} Using {{Recency Sampling}}}. In
  \bibinfo{booktitle}{\emph{Proc. {{RecSys}}}}. \bibinfo{pages}{81--91}.
\newblock


\bibitem[Petrov and Macdonald(2022b)]%
        {Bert4RecRepro}
\bibfield{author}{\bibinfo{person}{Aleksandr~V. Petrov} {and}
  \bibinfo{person}{Craig Macdonald}.} \bibinfo{year}{2022}\natexlab{b}.
\newblock \showarticletitle{A {{Systematic Review}} and {{Replicability Study}}
  of {{BERT4Rec}} for {{Sequential Recommendation}}}. In
  \bibinfo{booktitle}{\emph{Proc. {{RecSys}}}}. \bibinfo{pages}{436--447}.
\newblock


\bibitem[Petrov and Macdonald(2023a)]%
        {GPTRec}
\bibfield{author}{\bibinfo{person}{Aleksandr~V. Petrov} {and}
  \bibinfo{person}{Craig Macdonald}.} \bibinfo{year}{2023}\natexlab{a}.
\newblock \showarticletitle{Generative {{Sequential Recommendation}} with
  {{GPTRec}}}. In \bibinfo{booktitle}{\emph{Proc. {{Gen-IR}}@{{SIGIR}}}}.
\newblock


\bibitem[Petrov and Macdonald(2023b)]%
        {petrovGSASRecReducingOverconfidence2023}
\bibfield{author}{\bibinfo{person}{Aleksandr~V. Petrov} {and}
  \bibinfo{person}{Craig Macdonald}.} \bibinfo{year}{2023}\natexlab{b}.
\newblock \showarticletitle{{{gSASRec}}: {{Reducing Overconfidence}} in
  {{Sequential Recommendation Trained}} with {{Negative Sampling}}}. In
  \bibinfo{booktitle}{\emph{Proc. {{RecSys}}}}. \bibinfo{pages}{116--128}.
\newblock


\bibitem[Radford et~al\mbox{.}(2019)]%
        {gpt2}
\bibfield{author}{\bibinfo{person}{Alec Radford}, \bibinfo{person}{Jeffrey Wu},
  \bibinfo{person}{Rewon Child}, \bibinfo{person}{David Luan},
  \bibinfo{person}{Dario Amodei}, {and} \bibinfo{person}{Ilya Sutskever}.}
  \bibinfo{year}{2019}\natexlab{}.
\newblock \showarticletitle{Language {{Models}} Are {{Unsupervised Multitask
  Learners}}}.
\newblock \bibinfo{journal}{\emph{OpenAI blog}} (\bibinfo{year}{2019}).
\newblock


\bibitem[Raffel et~al\mbox{.}(2020)]%
        {raffelExploringLimitsTransfer2020}
\bibfield{author}{\bibinfo{person}{Colin Raffel}, \bibinfo{person}{Noam
  Shazeer}, \bibinfo{person}{Adam Roberts}, \bibinfo{person}{Katherine Lee},
  \bibinfo{person}{Sharan Narang}, \bibinfo{person}{Michael Matena},
  \bibinfo{person}{Yanqi Zhou}, \bibinfo{person}{Wei Li}, {and}
  \bibinfo{person}{Peter~J. Liu}.} \bibinfo{year}{2020}\natexlab{}.
\newblock \showarticletitle{Exploring the {{Limits}} of {{Transfer Learning}}
  with a {{Unified Text-to-Text Transformer}}}.
\newblock \bibinfo{journal}{\emph{Journal of Machine Learning Research}}
  \bibinfo{volume}{21}, \bibinfo{number}{140} (\bibinfo{year}{2020}),
  \bibinfo{pages}{1--67}.
\newblock
\showISSN{1533-7928}


\bibitem[Rendle et~al\mbox{.}(2009)]%
        {BPR}
\bibfield{author}{\bibinfo{person}{Steffen Rendle}, \bibinfo{person}{Christoph
  Freudenthaler}, \bibinfo{person}{Zeno Gantner}, {and} \bibinfo{person}{Lars
  {Schmidt-Thieme}}.} \bibinfo{year}{2009}\natexlab{}.
\newblock \showarticletitle{{{BPR}}: {{Bayesian}} Personalized Ranking from
  Implicit Feedback}. In \bibinfo{booktitle}{\emph{Proc. {{UAI}}}}.
  \bibinfo{pages}{452--461}.
\newblock


\bibitem[Schulman et~al\mbox{.}(2015)]%
        {schulmanTrustRegionPolicy2015a}
\bibfield{author}{\bibinfo{person}{John Schulman}, \bibinfo{person}{Sergey
  Levine}, \bibinfo{person}{Philipp Moritz}, \bibinfo{person}{Michael Jordan},
  {and} \bibinfo{person}{Pieter Abbeel}.} \bibinfo{year}{2015}\natexlab{}.
\newblock \showarticletitle{Trust Region Policy Optimization}. In
  \bibinfo{booktitle}{\emph{Proc. {{ICML}}}}. \bibinfo{pages}{1889--1897}.
\newblock


\bibitem[Schulman et~al\mbox{.}(2016)]%
        {schulmanHighDimensionalContinuousControl2016}
\bibfield{author}{\bibinfo{person}{John Schulman}, \bibinfo{person}{Philipp
  Moritz}, \bibinfo{person}{Sergey Levine}, \bibinfo{person}{Michael~I.
  Jordan}, {and} \bibinfo{person}{Pieter Abbeel}.}
  \bibinfo{year}{2016}\natexlab{}.
\newblock \showarticletitle{High-{{Dimensional Continuous Control Using
  Generalized Advantage Estimation}}}. In \bibinfo{booktitle}{\emph{Proc.
  {{ICLR}}}}.
\newblock


\bibitem[Schulman et~al\mbox{.}(2017)]%
        {schulmanProximalPolicyOptimization2017a}
\bibfield{author}{\bibinfo{person}{John Schulman}, \bibinfo{person}{Filip
  Wolski}, \bibinfo{person}{Prafulla Dhariwal}, \bibinfo{person}{Alec Radford},
  {and} \bibinfo{person}{Oleg Klimov}.} \bibinfo{year}{2017}\natexlab{}.
\newblock \bibinfo{title}{Proximal {{Policy Optimization Algorithms}}}.
\newblock
\newblock
\showeprint[arxiv]{1707.06347}~[cs]


\bibitem[Silver et~al\mbox{.}(2016)]%
        {silverMasteringGameGo2016}
\bibfield{author}{\bibinfo{person}{David Silver}, \bibinfo{person}{Aja Huang},
  \bibinfo{person}{Chris~J. Maddison}, \bibinfo{person}{Arthur Guez},
  \bibinfo{person}{Laurent Sifre}, \bibinfo{person}{George Van Den~Driessche},
  \bibinfo{person}{Julian Schrittwieser}, \bibinfo{person}{Ioannis Antonoglou},
  \bibinfo{person}{Veda Panneershelvam}, \bibinfo{person}{Marc Lanctot},
  \bibinfo{person}{Sander Dieleman}, \bibinfo{person}{Dominik Grewe},
  \bibinfo{person}{John Nham}, \bibinfo{person}{Nal Kalchbrenner},
  \bibinfo{person}{Ilya Sutskever}, \bibinfo{person}{Timothy Lillicrap},
  \bibinfo{person}{Madeleine Leach}, \bibinfo{person}{Koray Kavukcuoglu},
  \bibinfo{person}{Thore Graepel}, {and} \bibinfo{person}{Demis Hassabis}.}
  \bibinfo{year}{2016}\natexlab{}.
\newblock \showarticletitle{Mastering the Game of {{Go}} with Deep Neural
  Networks and Tree Search}.
\newblock \bibinfo{journal}{\emph{Nature}} \bibinfo{volume}{529},
  \bibinfo{number}{7587} (\bibinfo{year}{2016}), \bibinfo{pages}{484--489}.
\newblock


\bibitem[Stamenkovic et~al\mbox{.}(2022)]%
        {stamenkovicChoosingBestBoth2022}
\bibfield{author}{\bibinfo{person}{Dusan Stamenkovic},
  \bibinfo{person}{Alexandros Karatzoglou}, \bibinfo{person}{Ioannis Arapakis},
  \bibinfo{person}{Xin Xin}, {and} \bibinfo{person}{Kleomenis Katevas}.}
  \bibinfo{year}{2022}\natexlab{}.
\newblock \showarticletitle{Choosing the {{Best}} of {{Both Worlds}}:
  {{Diverse}} and {{Novel Recommendations}} through {{Multi-Objective
  Reinforcement Learning}}}. In \bibinfo{booktitle}{\emph{Proc. {{WSDM}}}}.
  \bibinfo{pages}{957--965}.
\newblock


\bibitem[Sun et~al\mbox{.}(2019)]%
        {BERT4Rec}
\bibfield{author}{\bibinfo{person}{Fei Sun}, \bibinfo{person}{Jun Liu},
  \bibinfo{person}{Jian Wu}, \bibinfo{person}{Changhua Pei},
  \bibinfo{person}{Xiao Lin}, \bibinfo{person}{Wenwu Ou}, {and}
  \bibinfo{person}{Peng Jiang}.} \bibinfo{year}{2019}\natexlab{}.
\newblock \showarticletitle{{{BERT4Rec}}: {{Sequential Recommendation}} with
  {{Bidirectional Encoder Representations}} from {{Transformer}}}. In
  \bibinfo{booktitle}{\emph{Proc. {{CIKM}}}}. \bibinfo{pages}{1441--1450}.
\newblock


\bibitem[Sunehag et~al\mbox{.}(2015)]%
        {sunehagDeepReinforcementLearning2015}
\bibfield{author}{\bibinfo{person}{Peter Sunehag}, \bibinfo{person}{Richard
  Evans}, \bibinfo{person}{Gabriel {Dulac-Arnold}}, \bibinfo{person}{Yori
  Zwols}, \bibinfo{person}{Daniel Visentin}, {and} \bibinfo{person}{Ben
  Coppin}.} \bibinfo{year}{2015}\natexlab{}.
\newblock \bibinfo{title}{Deep {{Reinforcement Learning}} with {{Attention}}
  for {{Slate Markov Decision Processes}} with {{High-Dimensional States}} and
  {{Actions}}}.
\newblock
\newblock
\showeprint{1512.01124}~[cs]


\bibitem[Sutton and Barto(2018)]%
        {suttonReinforcementLearningIntroduction2018}
\bibfield{author}{\bibinfo{person}{Richard~S. Sutton} {and}
  \bibinfo{person}{Andrew~G. Barto}.} \bibinfo{year}{2018}\natexlab{}.
\newblock \bibinfo{booktitle}{\emph{Reinforcement Learning: An Introduction}
  (\bibinfo{edition}{second edition} ed.)}.
\newblock \bibinfo{publisher}{The MIT Press}.
\newblock
\showLCCN{Q325.6 .R45 2018}


\bibitem[{TensorFlow Developers}(2023)]%
        {tensorflowdevelopersTensorFlow2023}
\bibfield{author}{\bibinfo{person}{{TensorFlow Developers}}.}
  \bibinfo{year}{2023}\natexlab{}.
\newblock \bibinfo{title}{{{TensorFlow}}}.
\newblock
\newblock
\urldef\tempurl%
\url{https://www.tensorflow.org/}
\showURL{%
\tempurl}


\bibitem[Vaswani et~al\mbox{.}(2017)]%
        {Transformer}
\bibfield{author}{\bibinfo{person}{Ashish Vaswani}, \bibinfo{person}{Noam
  Shazeer}, \bibinfo{person}{Niki Parmar}, \bibinfo{person}{Jakob Uszkoreit},
  \bibinfo{person}{Llion Jones}, \bibinfo{person}{Aidan~N Gomez},
  \bibinfo{person}{{\L}ukasz Kaiser}, {and} \bibinfo{person}{Illia
  Polosukhin}.} \bibinfo{year}{2017}\natexlab{}.
\newblock \showarticletitle{Attention Is {{All}} You {{Need}}}. In
  \bibinfo{booktitle}{\emph{Proc. {{NeurIPS}}}}.
\newblock


\bibitem[Wolf et~al\mbox{.}(2020)]%
        {wolfHuggingFaceTransformersStateoftheart2020}
\bibfield{author}{\bibinfo{person}{Thomas Wolf}, \bibinfo{person}{Lysandre
  Debut}, \bibinfo{person}{Victor Sanh}, \bibinfo{person}{Julien Chaumond},
  \bibinfo{person}{Clement Delangue}, \bibinfo{person}{Anthony Moi},
  \bibinfo{person}{Pierric Cistac}, \bibinfo{person}{Tim Rault},
  \bibinfo{person}{R{\'e}mi Louf}, \bibinfo{person}{Morgan Funtowicz},
  \bibinfo{person}{Joe Davison}, \bibinfo{person}{Sam Shleifer},
  \bibinfo{person}{Patrick {von Platen}}, \bibinfo{person}{Clara Ma},
  \bibinfo{person}{Yacine Jernite}, \bibinfo{person}{Julien Plu},
  \bibinfo{person}{Canwen Xu}, \bibinfo{person}{Teven~Le Scao},
  \bibinfo{person}{Sylvain Gugger}, \bibinfo{person}{Mariama Drame},
  \bibinfo{person}{Quentin Lhoest}, {and} \bibinfo{person}{Alexander~M. Rush}.}
  \bibinfo{year}{2020}\natexlab{}.
\newblock \bibinfo{title}{{{HuggingFace}}'s {{Transformers}}:
  {{State-of-the-art Natural Language Processing}}}.
\newblock
\newblock
\showeprint[arxiv]{1910.03771}~[cs]


\bibitem[Xin et~al\mbox{.}(2020)]%
        {xinSelfSupervisedReinforcementLearning2020}
\bibfield{author}{\bibinfo{person}{Xin Xin}, \bibinfo{person}{Alexandros
  Karatzoglou}, \bibinfo{person}{Ioannis Arapakis}, {and}
  \bibinfo{person}{Joemon~M. Jose}.} \bibinfo{year}{2020}\natexlab{}.
\newblock \showarticletitle{Self-{{Supervised Reinforcement Learning}} for
  {{Recommender Systems}}}. In \bibinfo{booktitle}{\emph{Proc. {{SIGIR}}}}.
  \bibinfo{pages}{931--940}.
\newblock


\bibitem[Zhang et~al\mbox{.}(2022)]%
        {zhangMultiTaskFusionReinforcement2022}
\bibfield{author}{\bibinfo{person}{Qihua Zhang}, \bibinfo{person}{Junning Liu},
  \bibinfo{person}{Yuzhuo Dai}, \bibinfo{person}{Yiyan Qi},
  \bibinfo{person}{Yifan Yuan}, \bibinfo{person}{Kunlun Zheng},
  \bibinfo{person}{Fan Huang}, {and} \bibinfo{person}{Xianfeng Tan}.}
  \bibinfo{year}{2022}\natexlab{}.
\newblock \showarticletitle{Multi-{{Task Fusion}} via {{Reinforcement
  Learning}} for {{Long-Term User Satisfaction}} in {{Recommender Systems}}}.
  In \bibinfo{booktitle}{\emph{Proc. {{KDD}}}}. \bibinfo{pages}{4510--4520}.
\newblock


\bibitem[Zhou et~al\mbox{.}(2010)]%
        {zhouSolvingApparentDiversityaccuracy2010}
\bibfield{author}{\bibinfo{person}{Tao Zhou}, \bibinfo{person}{Zolt{\'a}n
  Kuscsik}, \bibinfo{person}{Jian-Guo Liu}, \bibinfo{person}{Mat{\'u}{\v s}
  Medo}, \bibinfo{person}{Joseph~Rushton Wakeling}, {and}
  \bibinfo{person}{Yi-Cheng Zhang}.} \bibinfo{year}{2010}\natexlab{}.
\newblock \showarticletitle{Solving the Apparent Diversity-Accuracy Dilemma of
  Recommender Systems}.
\newblock \bibinfo{journal}{\emph{Proceedings of the National Academy of
  Sciences}} \bibinfo{volume}{107}, \bibinfo{number}{10}
  (\bibinfo{year}{2010}), \bibinfo{pages}{4511--4515}.
\newblock


\bibitem[Zou et~al\mbox{.}(2019)]%
        {zouReinforcementLearningOptimize2019}
\bibfield{author}{\bibinfo{person}{Lixin Zou}, \bibinfo{person}{Long Xia},
  \bibinfo{person}{Zhuoye Ding}, \bibinfo{person}{Jiaxing Song},
  \bibinfo{person}{Weidong Liu}, {and} \bibinfo{person}{Dawei Yin}.}
  \bibinfo{year}{2019}\natexlab{}.
\newblock \showarticletitle{Reinforcement {{Learning}} to {{Optimize Long-term
  User Engagement}} in {{Recommender Systems}}}. In
  \bibinfo{booktitle}{\emph{Proc. KDD}}. \bibinfo{pages}{2810--2818}.
\newblock


\end{thebibliography}
\FloatBarrier

\appendix
\section{Extra information on Reinforcement learning} \label{sec:rl}
\subsection{Basics of Reinforcement Learning}\label{ssec:rl_basics}

\sj{
In this section, we describe the basics of reinforcement learning necessary for describing our methodology.}
\looseness -1 \sasha{
    Reinforcement Learning~\cite{suttonReinforcementLearningIntroduction2018} (RL) is a class of \sj{machine learning} methods in which an \emph{agent} interacts with an \emph{environment} by performing \emph{actions}. The environment reacts to the actions by giving the agent \emph{rewards} and changing the agent's \emph{state}. The agent learns a \emph{policy} of selecting actions depending on the state that maximises the \sj{\emph{long-term reward} (reward accumulated over multiple steps)}.
}

\looseness -1 \sj{We now introduce some definitions and notations used across the paper to describe the Reinforcement Learning problem setup;  Typically, a Reinforcement Learning problem is formalised using the framework of \emph{Markov Decision Process} (MDP)~\cite[Ch.3]{suttonReinforcementLearningIntroduction2018}. While there are some variations in the definition of MDP in the literature, we use the definitions and notations from the TRPO paper~\cite{schulmanTrustRegionPolicy2015a} that largely influenced the Proximal Policy Algorithm~\cite{schulmanProximalPolicyOptimization2017a} -- the main RL algorithm we use in this paper.  In particular, a Markov Decision Process is defined as a tuple $MDP = \left<\mathcal{S}, \mathcal{A}, P, r, \rho_0, \gamma \right> \label{eq:mdp}$ 
where $S$ is a finite set of states, $\mathcal{A}$ is a finite set of actions, $P:\mathcal{S}\times \mathcal{A}\times \mathcal{S} \rightarrow \mathbb{R}$ is the distribution of transition probability, $r:\mathcal{S} \rightarrow \mathbb{R}$ is the reward function, $\rho_0: \mathcal{S} \rightarrow \mathbb{R}$ is the distribution of initial state $s_0$, $\gamma \in [0, 1]$ is the discount factor that defines how much we prefer immediate award over the award at the next step. %
}
\pageenlarge{3}
\sj{
    Given an $MDP$, the agent-environment interaction dynamics can be described as follows:
    \begin{enumerate}
        \item The initial state $s_0$ is drawn from the distribution $\rho_0$
        \item At each time step $i$, the agent observes the state $s_i \in \mathcal{S}$  and selects an action $a_i \in \mathcal{A}$
        \item The environment draws a new state $s_{i+1}$ from the transition probability distribution $P(s_{i+1}| s_i, a_i)$
        \item The agent receives a reward $r(s_{i+1})$
    \end{enumerate}
    The ultimate goal of the agent is this setup to maximise the long-term discounted reward $r_{0:\infty} = \sum_{i=0}^{\infty}\gamma^{i}r_i \label{eq:infinite_reward}$. 
    To achieve its goal, the agent follows a \emph{stochastic policy}, which is defined as a probability distribution $\pi: \mathcal{A} \times \mathcal{S} \rightarrow \mathbb{R}$. When the agent observes the state $s_i$, it selects the next action $a_i$ by drawing the action from the conditional distribution $\pi(a_i | s_i)$. In a typical Reinforcement Learning scenario, the agent also adjusts its policy when observing rewards and state transitions from the environment. In deep learning, the policy is typically modelled using a neural network $\pi_\Theta(a|s)$ where $\Theta$ is the set of network parameters. In that case, the policy is adjusted by changing the set of parameters $\Theta$. 
}

\looseness -1    Note that in the general case, the parameters of the MDP, such as the transition probability distribution $P(s_{i+1}| a_i, s_i)$ as well as the reward function $r(s)$ are not known to the agent (however, an agent may have \emph{some} prior knowledge about them); The agent has to \emph{learn} the structure of these parameters through the interactions with the environment and reflect this structure in the parameters of its policy.     While, generally speaking, the number of interactions that an agent performs with the environment may be infinite, in practice, in many situations, there is usually a \emph{terminal} state, after which the MDP stops. The process then may restart again. A single sequence of state actions that end in a terminal state is known as an \emph{episode}. The agent may use the knowledge obtained from interactions with the interactions in the previous episode to maximise the reward in the next episode. 

\sj{Reinforcement Learning is an active area of research, and there are many ways of learning optimal policies. For a comprehensive overview of the problem and the methodology, we refer to the classic book~\cite{suttonReinforcementLearningIntroduction2018}.  In our work, we use the \emph{Proximal Policy Optimisation} (PPO) algorithm as the main algorithm for reinforcement learning, because this algorithm has recently shown remarkable results in achieving goals similar to ours in the domain of Language Modelling. We describe the details of the PPO algorithm with application to our recommendation task in Section~\ref{ssec:ppo}.}

\sj{We now briefly describe how Reinforcement learning has been applied to Recommender Systems.}

\subsection{\sasha{Reinforcment Learning for RecSys}  \label{ssec:rl_background}}

    The use of RL is appealing for recommender systems because it can optimise for complex beyond-accuracy rewards \sjj{, which are not necessarily differentiable (e.g. diversity), } as well as optimise for long-term goals, which are hard to optimise using traditional supervised learning. Indeed, we are not the first to apply RL-based techniques for recommender systems~\cite{zouReinforcementLearningOptimize2019, zhangMultiTaskFusionReinforcement2022, stamenkovicChoosingBestBoth2022, deffayetGenerativeSlateRecommendation2023, afsarReinforcementLearningBased2023, baiModelbasedReinforcementLearning2019}. 
    
 Generally, existing approaches can be divided into two types based on what these methods use as the action space \sj{$\mathcal{A}$}. We now briefly describe these existing types. 
 
\pageenlarge{3}    
\paragraph{T1: Item-based methods}  \looseness -1 In this type, the methods (e.g.~\cite{stamenkovicChoosingBestBoth2022, xinSelfSupervisedReinforcementLearning2020})) use possible items $i \in I$ as the actions and make the recommendation by sorting the items according to the probability of selecting them as the next action (i.e. using the Top-K strategy). However, as we argue in Section~\ref{sec:intro}, it is hard to optimise the models that use the Top-K strategy for complex metrics, such as diversity, \sj{because these methods compute items' scores independently from each other.}  
    
    \paragraph{T2: Slate-based methods} \looseness -1 Methods (e.g.~\cite{deffayetGenerativeSlateRecommendation2023, sunehagDeepReinforcementLearning2015}) in this type use all possible \emph{slates} (full lists of recommended items $g \in \mathcal{G}$) as the action space and generates the slate using a single model inference. The problem with this group of approaches is that the space of all possible recommendation lists is very large (proportional \sdd{to $|I|^K$}, where $|I|$ is the number of possible items, and $k$ is the size of the slate). It is computationally infeasible to score all possible slates to select the best one. 
    Therefore, slate-based models usually require an existing source of high-quality slates that they can use as training data, which is not always available, or rely on some other heuristics, which limit the search space. Both \sdd{of} these groups of methods use the historical user-item interactions as the agent's state, using which the agent selects actions.

\paragraph{T3. Generative}
The method proposed in this paper \sdd{does not} fall into either of the existing types and therefore, we put it in a separate group. Indeed, similar to the methods from the slate-based method, it focuses on optimising listwise metrics; however, similar to the methods from the first group, the actions in our method correspond to items. We achieve \sdd{this} by changing the state space $\mathcal{S}$: in our case, the state includes not only historical user-item interactions but also a partially generated recommendations list. It allows the method to take into account items already recommended to the user, as well as plan for future actions ahead. In contrast to existing slate-based methods, our approach does not require existing sources of high-quality slates. Instead, it uses a 2-stage approach, where, in the first stage, the model learns to mimic the behaviour of the teacher model, and in the second stage, it aligns the model with the given effectiveness measure $R$. 

\looseness -1 The limitation of our method compared to the existing methods is that it requires inferring the recommendation model $K$ \sdd{to make a ranking of $K$ items}, whereas both Top-K methods and slate-based methods produce the list in just a single model inference; however, in practice, we find that our model is still able to generate recommendations reasonably fast: generating a recommendation on our hardware requires approximately 300 milliseconds, which is enough for \sj{many} practical applications. \sj{This is similar to interactive language model applications, such as ChatGPT, which generate a few tokens per second and still remain useful for a broad range of tasks.} 

\subsection{Details of the PPO Algorithm} \label{ssec:ppo}
\looseness -1 \sj{Proximal Policy Optimisation (PPO)} is a state-of-the-art algorithm for many Reinforcement Learning problems. As we describe in Section~\ref{ssec:rl_basics},  the goal of Reinforcement Learning is to train an \emph{agent} (a recommender system in our case) to perform \emph{actions} (choosing an item to recommend at position $i$) given a \emph{state} (a user with history $h$ and a partially generated list $g^{(i)} = [g_1, g_2, .. g_{i-1}]$) to optimise the long-term \emph{reward} (metric $R$ over the fully generated recommendation list $g$).  PPO assumes that the agent follows a stochastic \emph{policy $\pi_\Theta$} parametrised by a set of parameters $\Theta$: the agent selects actions from a probability distribution over the set of all possible actions, and PPO iteratively improves the policy by updating its parameters $\Theta$.
    
\sj{
    In our case, the policy defined by the GPTRec model itself defines the policy $\pi_{\Theta}$, and therefore PPO iteratively updates GPTRec's parameters to maximise effectiveness $R$. 
}

PPO follows the Actor-Critic optimisation scheme: it uses an auxiliary model, $V_\Psi$, parametrised by a set of parameters $\Psi$, which predicts the expected discounted long-term reward given the current state. The idea is to increase the probabilities of actions that increase our expected reward at each training iteration and decrease the probabilities of actions that decrease our expected reward. 
Therefore, in addition to training the main policy model, we train the value model:
\begin{align}
    V_\Psi(h, g^{(i)}) \approx \mathbb{E}_{\pi_{\Theta}} \left[\sum_{j \in 0 .. K-i-1} \gamma^{j}\mathrm{r}(h;g^{(i+j)}) \right] \label{eq:value}
\end{align}
here, $0 < \gamma \le 1$  is the discount hyperparameter, which controls how much the model prefers immediate reward over delayed reward, and $\mathbb{E}_{\pi_{\Theta}}$ means expectation over all possible generated recommendations with the GPTRec model. To optimise the value model, PPO uses a standard mean squared loss: 
\begin{align}
    \mathcal{L}_{MSE}(\Psi) = -\sum_{i=1}^K \left( V_\Psi(h, g^{(i)})  - \sum_{j=0}^{K-i-1}  \gamma^{j}\mathrm{r}(h;g^{(i+j)})\right)^2
\end{align}

\slw{
    PPO uses the value function to compute \emph{advantages}, which measures how much better or worse our model selected an item at each generation step compared to the expected value (according to metric $R$ and expected reward returned by $V$).
    PPO uses the \emph{Generalised Advantage Estimators} (GAE)~\cite{schulmanHighDimensionalContinuousControl2016}. 
    To compute the GAE advantages, we first compute the temporal difference (TD) residual. The TD residual,  denoted as $\delta_i$, at position  $i$ is given by:
    \begin{equation}
        \delta_i =\mathrm{r}(h, g^{(i + 1)}) + \gamma V_\Psi(h, g^{(i + 1)}) - V_\Psi(h, g^{(i)})
    \end{equation}
    where $\mathrm{r}(h, g^{(i + 1)})$ is the immediate reward at position $i+1$ (Equation~\eqref{eq:immediate_rewards}), and  $\gamma$ is the same discount factor as used in Equation~\eqref{eq:value}. 
    The GAE advantage fo position $i$, denoted as $A_i$, computes the advantage at each position $i$ using a decay factor $\lambda$ (which is also a hyperparameter that similarly to $\gamma$ controls the model focus on long/short term rewards) and TD residuals, is then  by:
    \begin{equation}
        A_i = \sum_{l=0}^{K - i - 1} (\gamma \lambda)^l \delta_{i+l} \label{eq:gae}
    \end{equation}
}
\pageenlarge{3}
\slw{
    To drive the model towards better rewards, PPO optimises the surrogate objective, which depends on the Ratio($\cdot$) function: 
    \begin{align}
        \text{Ratio}(\Theta, h, g^{(i)}) = \frac{\pi_\Theta(h,g^{(i-1)}, g_{i})}{\pi_{\Theta_{old}}(h,g^{(i-1)}, g_{i})} \label{eq:ratio}
    \end{align}
    where $\Theta_{old}$ is the set of model parameters used at the time of generating the recommendations list $g^{(i)}$. Note that PPO uses the same generated recommendations for multiple optimisation steps, and therefore,  $\Theta_{old}$ is not equal to $\Theta$, and overall, the proportion in Equation~\eqref{eq:ratio} is not equal to 1. 
    The PPO surrogate objective (also known as the CLIP objective) is then defined as:
}\slw{
    \begin{align}
        \mathcal{L}_{CLIP}(\Theta) &= -\sum_{i=1}^K \min \bigg(  \text{Ratio}(\Theta, h, g^{(i)}) A_i, \nonumber\\ & \text{clip}\left( \text{Ratio}(\Theta, h, g^{(i)}), 1-\epsilon, 1+\epsilon \right) A_i \bigg)
    \end{align}
    where the $\text{clip}(\cdot)$ function limits the minimum and maximum of the first argument, and $\epsilon$ is a hyperparameter that controls maximum divergence from old model parameters $\Theta_{old}$. Without clipping, the Ratio function (Equation~\eqref{eq:ratio}) is known to diverge from 1 very quickly; therefore, clipping is required to prevent \sjj{an} explosion of gradients of the CLIP objective.}
    
    Overall, the loss function optimised by PPO can be written as: 
    \begin{align}
        \mathcal{L}_{PPO}(\Psi, \Theta) =  \mathcal{L}_{MSE}(\Psi) + \mathcal{L}_{CLIP}(\Theta) \label{eq:ppo_loss}
    \end{align}
    The original PPO paper also includes an entropy bonus in the loss function (which encourages the model to perform exploration). However, in our initial experiments, we observed no improvement in the overall quality of GPTRec \sjj{when using an entropy bonus}; therefore, we omit it for training GPTRec. 
    Overall, PPO  can be summarised as repeating the following steps: 
        (Step 1) Sample a batch of user histories $h_1, h_2 ... h_B$; 
        (Step 2) for each history $h$ in the batch, generate recommendations $g$ using the current version of GPTRec $\pi_\Theta$. Note that at this step, we generate recommendations stochastically, drawing items from the distribution stochastically rather than selecting the item maximum score;
        (step 3) for each pair history-recommendations pair $\langle h, g \rangle$, compute advantages $A_1, A_2 \ldots, A_K$ using Equation \eqref{eq:gae};
        (step 4) perform multiple optimisation steps with respect to value function parameters $\Psi$ and GPTRec parameters $\Theta$, using gradient descent with $\mathcal{L}_{PPO}$ loss (Equation~\eqref{eq:ppo_loss}). 
    
This concludes the overview of the PPO algorithm for GPTRec optimisation. For more details, we refer to the original PPO and GAE papers~\cite{schulmanProximalPolicyOptimization2017a, schulmanHighDimensionalContinuousControl2016}.
\pageenlarge{3}
\vspace{-0.5\baselineskip}
\section{Sequence strucutre} \label{ssec:sequence_structure}
\www{Similar to the original GPT-2 model, GPTRec generates recommendations autoregressively using the Next-K recommendations strategy. This means model outputs at the $i$-th recommendations stage become model inputs at the $i+1$-th stage (alongside the original user-item interaction history). 
\sw{To help GPTRec's training,} we propose \sw{a input sequence structure that} consists of 4 parts:}
\begin{figure}[tb]
    \centering
    \vspace{-0.5\baselineskip}
    \resizebox{\linewidth}{!}{
    \includegraphics{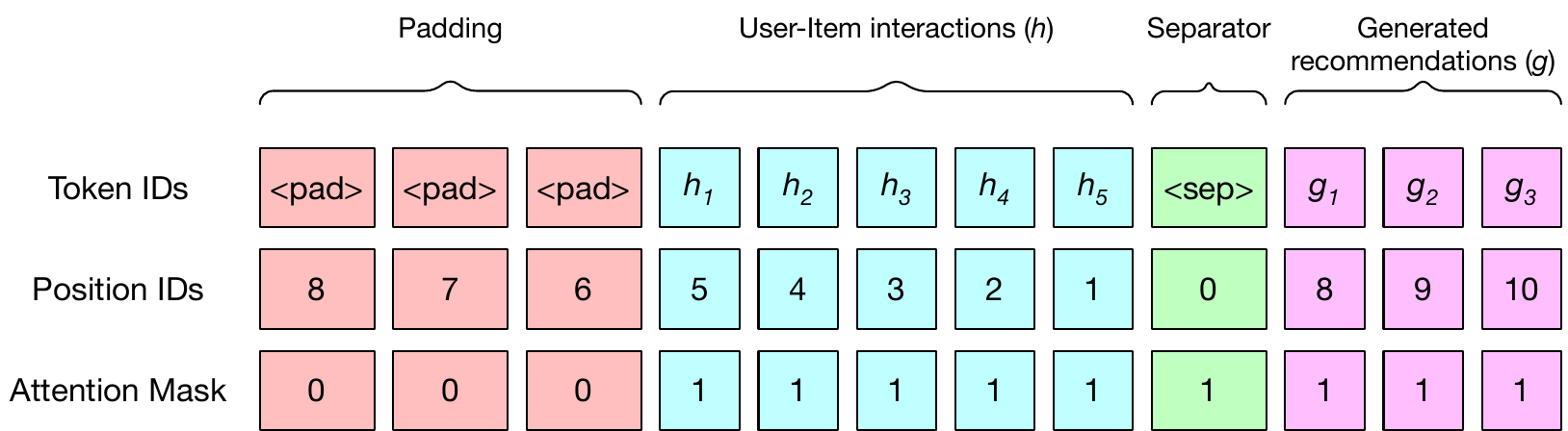}
    }
    \caption{\sw{Proposed input sequence structure}}
    \label{fig:sequence_structure}
    \vspace{-1\baselineskip}
\end{figure}

\begin{enumerate}
    \item optional padding to equalise lengths of all sequences;
    \item user-item historical interactions \sdd{$h=\{h_1, h_2.. h_n\}; h_i \in I$;
    \item separator symbol that mark the end of the "history"  part; 
    \item generated recommendations. $g=\{g_1, g_2.. g_k\}; g_i \in I$}
\end{enumerate}
\looseness -1 \www{Each position in the sequence consists of three integer numbers: (1) a token ID (the item ID, or one of two special tokens for padding and separator); (2) a position ID used by the Transformer architecture to preserve positional information; and (3) an attention mask, which tells GPTRec the positions in the sequence it can ignore.}

\looseness -1 \sdm{Initially, the sequence consists of the user history $h$ followed by a separator token}; then, at each next generation step $i$, the sequence shifts left, and a recommended item $g_i$ is added to the end of the sequence.  Because GPTRec shifts the same sequence multiple times during generation, the same position may have different semantics. For example,  the second last position belongs to the "history" part at the first generation step, after the first sequence shift separator moves to the second to last place, and at the third step, it is occupied by the first recommended item. 
Therefore, to help the model distinguish between different parts of the sequence, we adopt the following position ID scheme: \\
    $\bullet$  The separator token always has position ID 0; \\
    $\bullet$ The position IDs of historical user-item interactions are in reverse chronological order. This way, the last action (which is the most important for recommendation~\cite{PetrovRSS22}) always has position ID 1; the second last has position ID 2, etc.;\\
     $\bullet$ \sdm{The position IDs associated with recommended items are in ascending order starting from position ID $n+1$ where $n$ is the maximum length of the user-item interactions history; i.e. first recommended item has position ID $n+1$, second item $n+2$ etc.}  This way, each position in the recommendation is associated with a specific position ID that does not change during the generating process; \\
    $\bullet$ The model ignores item padding tokens, so their position IDs can be anything.

\FloatBarrier

\end{document}